\documentclass{article}
\usepackage{graphicx}
\usepackage{natbib}
\usepackage[margin=1in]{geometry}
\usepackage{authblk}

\usepackage{mathrsfs}
\usepackage{bm}
\usepackage{subcaption}

\usepackage[utf8]{inputenc}
\usepackage[T1]{fontenc}
\usepackage{amsmath}

\graphicspath{{Figures/}}

\def\be{{\bf E}}
\def\bb{{\bf B}}

\def\bve{{\bf V}_e}
\def\bvi{{\bf V}_i}
\def\ih{\hat{\bf x}}
\def\jh{\hat{\bf y}}
\def\kh{\hat{\bf z}}
\def\eps{\varepsilon}

\newcommand{\pd}[2]{\frac{\partial #1}{\partial #2}}

\title{The Nonlinear Evolution of Whistler-Mode Chorus Revisited: Modulation Instability as the Source of Tones}

\author{Daniel J. Ratliff$^{1}$ and Oliver Allanson$^{2,3,4}$}

\date{%
    $^1${\it  Department of Mathematics, Physic and Electrical Engineering, Northumbria University}\\[1ex]%
    $^2${\it Space Environment and Radio Engineering, School of Engineering, University of Birmingham, Birmingham, B15 2TT, UK}\\[1ex]%
       $^3${\it Department of Earth \& Environmental Sciences, University of Exeter, Penryn, TR10 9FE, UK} \\[1ex]%
          $^4${\it Department of Mathematics \& Statistics, University of Exeter, Exeter, EX4 4QF,  UK}\\[2ex]%
    \today
}

\begin{document}

\maketitle

\begin{abstract}
    We review the modulation stability of parallel propagating/field aligned Whistler Mode Chorus waves propagating in a warm plasma from a formal perspective with a focus on wave-particle interactions. The modulation instability criteria is characterised by a curvature of the dispersion relation for Whistler mode waves and a condition on the ratio between the group velocity $c_g$ and the electron sound speed $c_{s,e}$. We also demonstrate the in order to investigate the spatiotemporal evolution of the envelope and the formation of packets, one necessarily needs to account for the motion of ions within the system, leading to an ionic influence on the modulation instability threshold determined by the ion fraction of the plasma. Finally, we demonstrate that chirping may be captured when higher order effects are included within the spatiotemporal evolution of the amplitude. This yields not only an explicit expression for the sweep rate but identifies a possible origin for the power band gap that occurs at half the electron gyrofrequency. Numerical validation demonstrates that the interaction between wave packets is a source for the emergence of tones observed within mission data, and such interactions may be a major source of the electron energisation which Whistler-Mode chorus are responsible for.
\end{abstract}

\section{Introduction}

Whistler-mode chorus (WMC) waves play a significant role in determining energetic electron dynamics within terrestrial and magnetospheric plasmas \citep{Horne-2005Nature, Thorne-2010Nature, Artemyev-2016, Woodfield-2018}. WMC are one particular manifestation of the so-called `whistler-mode' electromagnetic/plasma wave \citep{Stix-1992}, and are particularly noteworthy for their role in rapid electron energisation and pitch-angle scattering \citep{Bortnik-2008GRL, Omura-2008, Albert-2010, Artemyev-2018, Zhang-2022}. Discussions of the role of WMC as one driver among many (within the general context of energetic charged particle dynamics in the inner magnetosphere) can be found in e.g. \citet{Green-2004, Thorne-2010, Bortnik-2016, Li(Wen)-2019, Lejosne-2022}. In perhaps overly simplistic terms, one can consider two main and contrasting challenges to achieving a full understanding of the role of WMC in magnetospheric plasma dynamics. While they are somewhat contrasting, both of these challenges are fundamentally united by the critical role that is played by wave-particle interactions \citep{Brice-1964, Kennel-1966b, Tsurutani-1997, Summers-1998}. 


One challenge is to determine the direct impact of WMC on electrons that would otherwise evolve adiabatically as a geomagnetically trapped particle \citep{Shklyar-2009, Albert-2022}. 
One of the most impressive manifestations of this approach (`wave effects on particles only') is the application of the resonant diffusion limit of the quasilinear theory (e.g. \citet{Kennel-1966, Summers-2005, Allanson-2022}) to global-scale numerical modelling of the terrestrial and planetary radiation belt populations using Fokker-Planck radiation-belt models (e.g. \citet{Li-2014, Glauert-2018, Wang-2020, Allison-2021}. One of the numerous outstanding problems in this area is to understand and incorporate the role of so-called `nonlinear wave-particle interactions' \citep{Artemyev-2021, Artemyev-2022c}, with WMC playing a very important role.

Another challenge is to instead try to solve for one or more of the generation, interaction and subsequent evolution of the WMC wave modes, as a function of e.g. a given initial plasma condition, and perhaps with some external driving or particle sources/injections. Studies of this nature try to understand the evolution of both the wave amplitude (i.e. amplitude amplification and modulation) and the structure in frequency space (i.e. either rising- or falling- tones, or even more exotic forms such as `hooks'). We should point out that the most general definition of WMC includes a variety of spectral forms, including comparatively structureless emissions (sometimes known as `hiss-like chorus', e.g. see \citet{Tsurutani-1974, Li-2012, Tsurutani-2013, Gao-2014, Shumko-2018}), as well as the more well-known and coherent/structured/`chirping' rising- and falling-tones \citep{Burtis-1976, Koons-1990, Li-2011b, Li-2012, Taubenschuss-2014, Santolik-2014, Teng-2019}. Approaches of this kind usually necessitate some form of `self-consistent approach', in which one ultimately solves some variation of the Vlasov-Maxwell system \citep{Schindlerbook} given a number of constraints.  Therefore one is likely solving first for the influence of unstable particle distributions on waves \citep{Gary-1993}, and possibly also for subsequent resulting turbulence induced by wave-particle interactions and/or wave-wave interactions \citep{Kadomstevbook, Sagdeev-1969}. 

A number of important open questions remain regarding both of these challenges, and of course the separation into these two contrasting approaches (which for the purposes of this discussion have been crudely polarised as `wave effect on particle only' and `wave evolution only') is an approximation to the complex, dynamic and networked energy pathways of the inner magnetospheric plasma \citep{Jaynes-2015, Li(Wen)-2019, Ripoll-2020, Koskinen-book}. 

There have been a number of thorough recent reviews and discussions of WMC generation and evolution \citep{g19, t20, t21, z21, o21} and so we do not do a complete literature review, instead directing the reader to those references and therein. It suffices to say that the inhomogeneity of the background magnetic field is frequently invoked to play a key role in the chirping mechanism for WMC \citep{Helliwell-1967, Sudan-1971, Nunn-1974, Vomvoridis-1982, Trakhtengerts-1995, Omura-2008, t21}, facilitating resonant particle trapping, bunching, and the formation of `resonant currents'. However there are some proposed mechanisms that do not rely upon the inhomogeneous background to drive the chirping behaviour \citep{z21, Zonca-2022}, and in particular, we note recent particle-in-cell numerical experiments that demonstrate chirping behaviour within the context of a uniform background magnetic field \citep{Wu-2020}. 

The most significant contribution of this work is to demonstrate the key role that the ponderomotive force can play in driving chirping behaviour in WMC within the context of a homogeneous background field. This phenomena could now be considered in addition to other aforementioned mechanisms.



 In this paper we simultaneously consider the role of wave-particle interactions both on the evolution of WMC and on the particle populations themselves due to ponderomotive forces. We will do so via the theory of modulations and weakly nonlinear theories, through which the evolution of the wave amplitude is coupled to the variations of number density, to observe how the interplay between the two manifests at the onset of nonlinear effects. Ultimately this will build upon the ideas introduced in previous theoretical treatments, and crucially those introduced by \cite{o08,o21} which investigated the evolution of WMC given a background particle population, now accounting for the simultaneous evolution of these species alongside the wave motion. However, we reiterate that the derivations presented in this work are limited to the case in which the background magnetic field is infinite and uniform. Therefore, in the context of e.g. the Earth's radiation belts, this implies that the analysis applies close to the geomagnetic equator. 


The approach that we take is facilitated by several observations. The first is that coherent WMC waves are known to be narrow banded, with the bandwidth being approximately $10\%$ of the local gyrofrequency~\citep{santolik-2003,santolik-2008}, meaning that one may restrict the study of the dynamics to a single wave mode. Further, the asymptotic picture is simplified further by the observation that the majority of nonlinear generation processes and amplification events occur near-equatorially in the magnetosphere and confined to very limited latitudes~\citep{lauben-2002,ledocq-1998,Meredith-2020}, meaning it is not unreasonable to restrict ourselves to a fluid system in cartesian co-ordinates without latitude considerations and that we are within the remit of weak curvilinear magnetic field effects. Finally, by considering the field-aligned case the reduction procedure is much more straightforward owing to the fact that the Lorentz force ${\bf v}\times {\bf B}$ no longer contributes to the nonlinear effects. This is due to the fact that when the wave and the velocity field are oriented in the same direction, the Lorentz force does not generate anything beyond linear terms in an asymptotic theory. A consequence of this is that higher harmonics, that is non-zero integer frequency  multiples of the carrier wave (e.g. $\pm 2 \omega,\,\pm 3 \omega,\,\ldots$), do not contribute to the wave motion. Instead, the wave evolution consists of simply the carrier wave interacting with the particle populations, and it is solely the wave-particle interaction via the ponderomotive forces that drives the nonlinearity observed in the wave's evolution.

With these simplifying factors considered, we are able to explore the emergence of wave packets in WMC using two main nonlinear approaches. The first of these is via classical modulation theory~\citep{wlnlw}, which postulates that the parameters of the wave such as the amplitude, frequency and wavenumber all evolve slowly over the course of many wave periods in a similar fashion to classical WKB theory. Such approaches have been successfully used within space plasmas~\citep{m76,gp77,m86,eliasson-2005,tracy-2014,o08,o21}, with the closest related work to this paper considering WMC to deduce the modulation stability of these waves~\citep{Tam-1969}, but with wave-particle interactions not fully accounted for and by use of a simplified version of of the dispersion relation. When a similar approach is also utilised for the electron beam velocity and number density with ponderomotive effects accounted for, we obtain our first insight into the ponderomotive-driven wave-particle interactions influencing WMC modulation. It is these interactions influencing the emergence of wave packets through electron-acoustic effects. Ultimately this builds upon the ideas introduced in theoretical treatments of the second approach, crucially those introduced by \cite{o08,o21} which investigated the evolution of WMC given a background particle population, and now account for the simultaneous evolution of these species alongside the wave motion.

The second approach is to undertake a formal multiple scales analysis to derive a direct evolution equation for the spatiotemporal evolution of WMC amplitude, taking the form of the Nonlinear Schr\"odinger (NLS) equation. This equation has emerged from heuristic arguments in prior works~\citep{karpman-1977,stenflo-1986}, with the key work of \cite{krafft-2018} highlighting its emergence within a wave-particle interaction framework. However, one of the results of this paper is to demonstrate that a formal asymptotic procedure reveals that one should be cautious using such approaches due to an inconsistency that emerges in the induction equation, which is overlooked in previous approaches. As a consequence it highlights that the motion of ions, and not just a population that ensures neutrality, must necessarily be considered to resolve this inconsistency, and in doing so we find that these wave-particle interactions are augmented. Our paper demonstrates that although some of the qualitative conclusions of \cite{krafft-2018} are the same, namely that the curvature of dispersion plays a role in the formation of solitons, the ion number fraction plays a critical role in whether WMC elements are to be observed. Therefore one of the main conclusions of this work is the statement that ion motion cannot be neglected in such problems.

With the understanding that arises from the above theoretical approaches, we are then able to augment these ideas within this paper by capturing WMC chirp, the mechanism behind rising and falling tones and one of the most intriguing features of WMC. This phenomenon presents itself as significant repetitive sweep of the dominant frequency peak of the wave. It is known that the NLS equation does not admit such behaviours and requires higher order effects to be introduced to account for this behaviour. This is precisely what we obtain as part of this work,  extending the multiple scales analysis we demonstrate from a formal perspective that this process arises from the wave-particle interactions. A surprising consequence of this is that the extended model may provide an explanation for the observed band gap in WMC waves at half the gyrofrequency \citep{Fu-2014,Gao-2019,Li-2011,Chen-2022} as the terms responsible for frequency variations vanish at precisely this frequency. As such, the theory suggests that sweep rates decrease for waves which approach the band gap before arresting completely. Overall, this analysis provides expressions for the sweep rate of a single WMC element, which  demonstrates that within isolation a WMC element cannot produce monotonic tones with a net change in the frequency of the wave. Instead, it suggests that the chirping behaviour observed originates from the interaction of several WMC elements, which we corroborate with numerical experiments. These produce repetitive WMC tones, and the space-time series demonstrates that their interaction also generates pulsations in the wave envelope which manifest in the particle dynamics as amplifications in the energy density. This, we speculate, may shed light on which stage of their propagation WMC might be energising the electron populations they trap during transit.

The outline of this paper is as follows. We begin with a review of the modulation instability of parallel propagating WMC in \S \ref{sec:Mod1}, a process entirely driven by the ponderomotive wave-particle interactions, outlining where such waves become unstable and form subpacket structures. Subsequently, we derive evolution equations for the wave envelope of WMC in \S \ref{sec:WNT}, leading to a Nonlinear Schr\"odinger (NLS) equation and revealing that ions play a crucial role in the envelope dynamics and alter the expected modulational stability transition. Owing to a lack of chirping behaviour, we add correction terms to the NLS that capture such effects in \S \ref{sec:chirp}, leading to explicit expressions for chirping within a single element. Finally, we use numerical simulations of this model in \S \ref{sec:numerics} to demonstrate how the interactions between wave packets is the main driver of the chirping seen within the mission data. Concluding remarks are given in \S \ref{sec:conc}.

\section{Review: Modulation Instability of Whistler Mode Waves in Electron-only Plasmas}
\label{sec:Mod1}
To understand the formation of WMC wave packets, we must first analyse the necessary conditions that permit their formation. This is done from the viewpoint of modulation instability, the process under which uniform wavetrains destabilise and undergo amplitude modulations, ultimately to form several wave elements as wave energy clusters within packets~\citep{Ablowitz-1981,Billingham-2000,wlnlw,Treumann-2001,cipp}. The approach to identify this instability is to derive quasilinear modulation equations governing the slow evolution of wave, typically amplitude and wavenumber,  and determine when this system possesses complex eigenvalues. Quasilinear modulation equations have been derived previously for the wave without particle interaction effects, with notable works relevant to our approach including \citep{Tam-1969,Omura-2008,Omura-2021}, but a key extension of this work will be to introduce modulation equations governing the electron number density and beam velocity. Such effects make a significant difference to the stability transition of the wave and thus it is pertinent to include such evolution simultaneously with the electromagnetic wave.

Throughout this paper, we will be considering collisonless, warm, isothermal plasmas from a fluid description. In the first instance, we will be considering a non-relativistic plasma purely comprising of electrons, neglecting any ion influences for the moment but we note these will be accounted for in later sections. Thus, we will be concerned with the following equations of motion:
\begin{gather}
\label{MHD-1}
    \nabla \times \be = -\pd{\bb}{t}\,,\\
    \nabla \times \bb  = \mu_0 q n \bve+\frac{1}{c^2}\pd{\be}{t}\,,\\
    \pd{\bve}{t}+(\bve\cdot \nabla)\bve +\frac{c_s^2}{n}\nabla n= \frac{q}{m}\left(\be+\bve \times \bb\right)+{\bf F}_P\,,\\
    \pd{n}{t}+\nabla \cdot (n\bve) = 0
\end{gather}
In the above, ${\bf B}$, ${\bf E}$, ${\bf V}$ and $n$ represents the magnetic field, electric field, electron velocity field and electron number density respectively. The parameters $q = -e$ and $m$ represent the charge and mass of an electron, $\mu_0$ is the magnetic permittivity constant, $c$ is the speed of light and $c_s^2 = k_BT/m$ is the speed of sound for the purely electron plasma. 
The ponderomotive force ${\bf F}_P$ acting on each electron is given by (see, for example,~\citep{cipp,nipt,lm83})
\[
{\bf F}_P = -\frac{\omega_{pe}^2}{2\mu c^2 \omega^2}\nabla \big( \langle |{\bf E}|^2\rangle\big)\,,
\]
where $\omega_{pe}$ is the electron plasma frequency and $\omega$ is the frequency of the electromagnetic wave.
In essence, this force determines the mean drift of electrons over rapid gyrofrequency oscillations due to amplitude modulations emerging within the monochromatic wavetrain. There are a number of choices one can make for this force depending on the plasma environment \citep{treumann-1997,krafft-2018}, but in this body of work we consider the simplest such force corresponding to the lowest order ponderomotive effect. This is to illustrate that the presence of such a force, even in its most rudimentary form, is crucial for wave-particle interactions and as a consequence the formation of chorus wave packets and elements.

The framework in which this will be achieved is through the use of WKB theory, or equivalently Whitham modulation theory. The starting point for this will be to consider the following Stokes wave ansatz for a wave-particle solution, representing a parallel propagating right polarised wave in the presence of a uniform magnetic field with strength $B_0$:
\[
\begin{split}
\bb &= B_0\kh+(\ih-i\jh)B_We^{i\theta}\\[3mm]
\be &= \alpha_1(\ih-i\jh)B_We^{i\theta}\\[3mm]
\bve &=v_{||}\kh + \alpha_2(\ih-i\jh)B_We^{i\theta}+c.c.+V\kh\\[3mm]
n &= n_0+N\,, \qquad \qquad \theta = \omega t - k z\,,
\end{split}
\]
where $z$ is in the direction of $\kh$. The parameters $v_{||}$ and $n_0$ represent the constant parallel velocity and reference electron number density respectively. The wave amplitude $B_W$, mean velocity perturbation $V$ and number density perturbation $N$ are initially assumed to be constant but small, so that $|B_W|\ll 1$. Typically one is able to characterise this smallness by comparing linear and leading order nonlinear terms, which for WMC is achieved by comparing the convective term with the Lorentz force in the momentum equation:
\[
 \left\vert\frac{B_W}{B_0}\right\vert\sin \phi \ll \left \vert\frac{\omega -v_{||}k}{\Omega}\right \vert
\]
where $\phi$ is the wave normal angle, that is the angle between the wave and background magnetic field.
It follows that this ordering of magnitude is trivially satisfied for any choice of the system parameters for parallel propagating WMC. Thus there is considerable freedom regarding the magnitude of waves this theory can consider, but must still be small enough to separate linear and nonlinear scales.
The expansion procedure requires that $N,\,V = \mathcal{O}(|B_W|^2)$, owing to the fact that these oscillation-free terms must balance the oscillation-free terms generated by the ponderomotive force.

The approach is to substitute this ansatz into the governing equations (\ref{MHD-1}) and consider terms up to $\mathcal{O}(|B_W|^3)$. The details of this calculation can be found within appendix \ref{app:WMT}, but we summarise the key elements of the approach here. The carrier wave terms, when substituted into the governing equations, generate the dispersion relation
\[
D(\omega, k,v_{||}) = \frac{1}{c^2}\bigg[(\omega-v_{||}k-\Omega_e)(c^2k^2-\omega^2)+\omega_{pe}^2(\omega-v_{||}k)\bigg]\,,
\]
which vanishes whenever the frequency $\omega = \omega_0(k,v_{||})$ satisfies the typical Whistler-mode dispersion curve
\[
c^2k^2 = \omega_0^2-\frac{\omega_{pe}^2(\omega_0-v_{||}k)}{\omega_0-v_{||}k-\Omega_e}\,.
\]
By continuing the analysis to higher powers of the amplitude to include amplitude-dependent weakly nonlinear effects, one finds 
the result
\begin{equation}\label{WKB-eqtns}
\begin{split}
D B_W+\frac{\omega_{pe}^2}{c^2n_0}\frac{(\omega-v_{||}k)^2-\Omega_e \omega}{\omega-v_{||}k-\Omega_e}NB_W = 0\,,\\
\frac{c_s^2-v_{||}^2}{n_0}N+\frac{\omega_{pe}^2}{\mu c^2 k^2} |B_W|^2 = \gamma -\frac{v_{||}^2}{2}-c_s^2\ln n_0\,.
\end{split}
\end{equation}
which can be used to extract the nonlinear dispersion relations
\begin{equation}\label{nl-disp}
\begin{split}
\omega &= \omega_0(k,v_{||}) +\frac{\omega_{pe}^2\big((\omega-v_{||}k)^2-\Omega_e \omega\big) }{\omega_{pe}^2\Omega_e+2\omega(\omega-v_{||}k-\Omega_e)^2}\frac{N}{n_0}\,, \\[3mm]
\gamma &= \frac{v_{||}^2}{2}+c_s^2\ln n_0+\frac{c_s^2-v_{||}^2}{n_0}N+\frac{\omega_{pe}^2}{\mu c^2 k^2} |B_W|^2\,.
\end{split}
\end{equation}
These are denoted as nonlinear dispersion relations due to the presence of the wave amplitude, mean beam velocity and number density variation as corrections to the linear dispersion.
The system (\ref{WKB-eqtns}) can be thought of as that obtained by \cite{Omura-2008,Omura-2021} but accounts for higher order nonlinear effects and inherent wave-particle interactions due to ponderomotive effects. In theory, the two approaches could be combined to augment the existing theory of the previous references, but this is not the focus of this work. Instead, to complete our analysis here, we demonstrate the above system can be cast in variational form, to make the subsequent analysis closer to classical wave modulation theory~\citep{wlnlw}. This is done by introducing %
\[
\mathscr{B} = \frac{k^2}{n_0}\frac{(\omega-v_{||}k)^2-\Omega_e \omega}{\omega-v_{||}k-\Omega_e}|B_W|^2\,,
\]
with the factor nonvanishing for Whistler waves, allowing the system can be written as
\[
\begin{split}
D +\frac{\omega_{pe}^2}{c^2n_0}\frac{(\omega-v_{||}k)^2-\Omega_e \omega}{\omega-v_{||}k-\Omega_e}N\equiv  D+QN = 0\,,\\
\frac{c_s^2-v_{||}^2}{n_0}N+Q\mathscr{B} = \gamma -\frac{v_{||}^2}{2}-c_s^2\ln n_0\,.
\end{split}
\]
The above system of equations is generated by the $\mathscr{B}$ and $N$ variations of Lagrangian density
\begin{equation}\label{Lag-MHD1}
\begin{split}
\mathscr{L} =& D\mathscr{B}+Q\mathscr{B}N+c_s^2N\ln n_0+\bigg(\frac{v_{||}^2}{2}-\gamma\bigg)(N+n_0)+\frac{1}{2}\frac{c_s^2-v_{||}^2}{n_0}N^2\,, \\[3mm]
&{\rm with} \qquad Q = \frac{\omega_{pe}^2}{c^2n_0}\frac{(\omega-v_{||}k)^2-\Omega_e \omega}{\omega-v_{||}k-\Omega_e}\,.
\end{split}
\end{equation}
This Lagrangian can be thought of as that which is averaged over one period of the Whistler mode wave. It is from this Lagrangian density that we will derive the conditions for the Whistler wave to undergo a modulational instability, associated with wave packet generation. Ultimately this will signpost this criteria in the electron case, which we will develop within the more applicable but involved ion-electron plasma case. 

\subsection{Modulation Instability}
\label{sec:MI-electron}
We may now study the Lagrangian (\ref{Lag-MHD1}) to determine when the parallel Whistler wave is expected to be stable and remain close to monochromatic. If it is not, it is expected to form packets (also known as elements) and as a result generate larger amplitude events. We do so by appealing to Whitham modulation theory\citep{w70,wlnlw} which has in the past been utilised in plasma contexts for the same purpose (see \citep{m76,gp77,m86} for some key examples). Notably we will be augmenting the existing literature on plasma modulations to account for wave-particle interaction effects, requiring the consideration of an additional phase for the background velocity. The crux of the approach is to introduce the `rapid' wave phase and velocity potential,
\[
\theta = \eps^{-1}\Theta(Z,T)\,, \quad \phi = \eps^{-1} \Phi(Z,T)\,,\quad Z = \eps z\,,\quad T = \eps t\,,\quad \eps \ll 1\,.
\]
This allows the wave parameters $k,\,\omega,\,v_{||}$ and $\gamma$ to vary slowly in space and time:
\[
k = -\Theta_Z,\,\quad \omega = \Theta_T,\, \quad v_{||}  = \phi_Z,\,\quad \gamma = -\phi_T\,\
\]
where subscripts denote partial derivatives with respect to the subscripted variable.
This admits the phase consistency conditions
\[
k_T+\omega_Z = 0\,, \quad (v_{||})_T+\gamma_Z = 0\,.
\]
We can replace $\omega$ and $\gamma$ using (\ref{nl-disp}) to obtain the first two modulation equations
\[
\begin{split}
k_T+\left(\omega_0(k,v_{||}) +\frac{\omega_{pe}^2\big((\omega-v_{||}k)^2-\Omega_e \omega\big) }{\omega_{pe}^2\Omega_e+2\omega(\omega-v_{||}k-\Omega_e)^2}\frac{N}{n_0}\right)_Z = 0\\
(v_{||})_T+\left(\frac{v_{||}^2}{2}+c_s^2\ln n_0+\frac{c_s^2-v_{||}^2}{n_0}N+\frac{\omega_{pe}^2}{\mu c^2 k^2} |B_W|^2\right)_Z = 0\,.
\end{split}
\]
The first of these equations is referred to as the conservation of waves, whereas the second is the classical conservation of momentum for the plasma in the presence of the wave.
The subsequent two equations, which close the system, come from taking the $\Theta$ and $\Phi$ variations of $\mathscr{L}$, which when simplified give
\[
\begin{split}
\mathscr{B}_T+\bigg(c_g \mathscr{B}-\frac{Q_k}{D_\omega}\mathscr{B}N\bigg)_Z = 0\,,\\
N_T+\bigg(v_{||}(N+n_0)-\frac{v_{||}}{n_0}N^2+Q_{v{||}}N\mathscr{B}\bigg)_Z = 0\,.
\end{split}
\]
These equations represent the conservation of wave action and the conservation of mass respectively.
These 4 equations can then be written in quasilinear form~\citep{wlnlw}:
\begin{equation}\label{mod-system}
\begin{split}
&{\bf U}_T+{\bf A}({\bf U}){\bf U}_Z = 0\,, \\[5mm]
{\rm with} \quad {\bf U} = \begin{pmatrix}k\\[2mm] \mathscr{B} \\[2mm] v_{||}\\[2mm] N\end{pmatrix}\,, \quad 
{\bf A} = &\begin{pmatrix}
c_g-\big(\frac{Q}{D_\omega}\big)_k N&0&\frac{\partial \omega_0}{\partial v_{||}}-\big(\frac{Q}{D_\omega}\big)_{v_{||}}N&-\frac{Q}{D_\omega}\\[2mm]
\omega_0''\mathscr{B}&c_g-\frac{Q_k}{D_\omega}N&(c_g)_{v_{||}}&-\frac{Q_k}{D_\omega}\mathscr{B}\\[2mm]
Q_k\mathscr{B}&Q&0&\frac{c_s^2}{n_0}\\[2mm]
0&Q_{v_{||}}N&n_0+N-\frac{N^2}{n_0}&Q_{v_{||}}\mathscr{B}
\end{pmatrix}
\end{split}
\end{equation}
This quasilinear system of equations encapsulate the wave-particle interactions for a single wave - the second equation determines how the wave amplitude is modified by the presence of variations to the number density, whilst simultaneously the fourth equation describes how the local number density is altered due to the wave. The remaining equations, the first and third, dictate how the local frequencies and velocity field responds to the wave-particle interaction respectively and thus alter the properties of the wave propagation.

\begin{figure}[ht]
    \centering
    \begin{subfigure}{0.4\textwidth}
    \includegraphics[width=\textwidth]{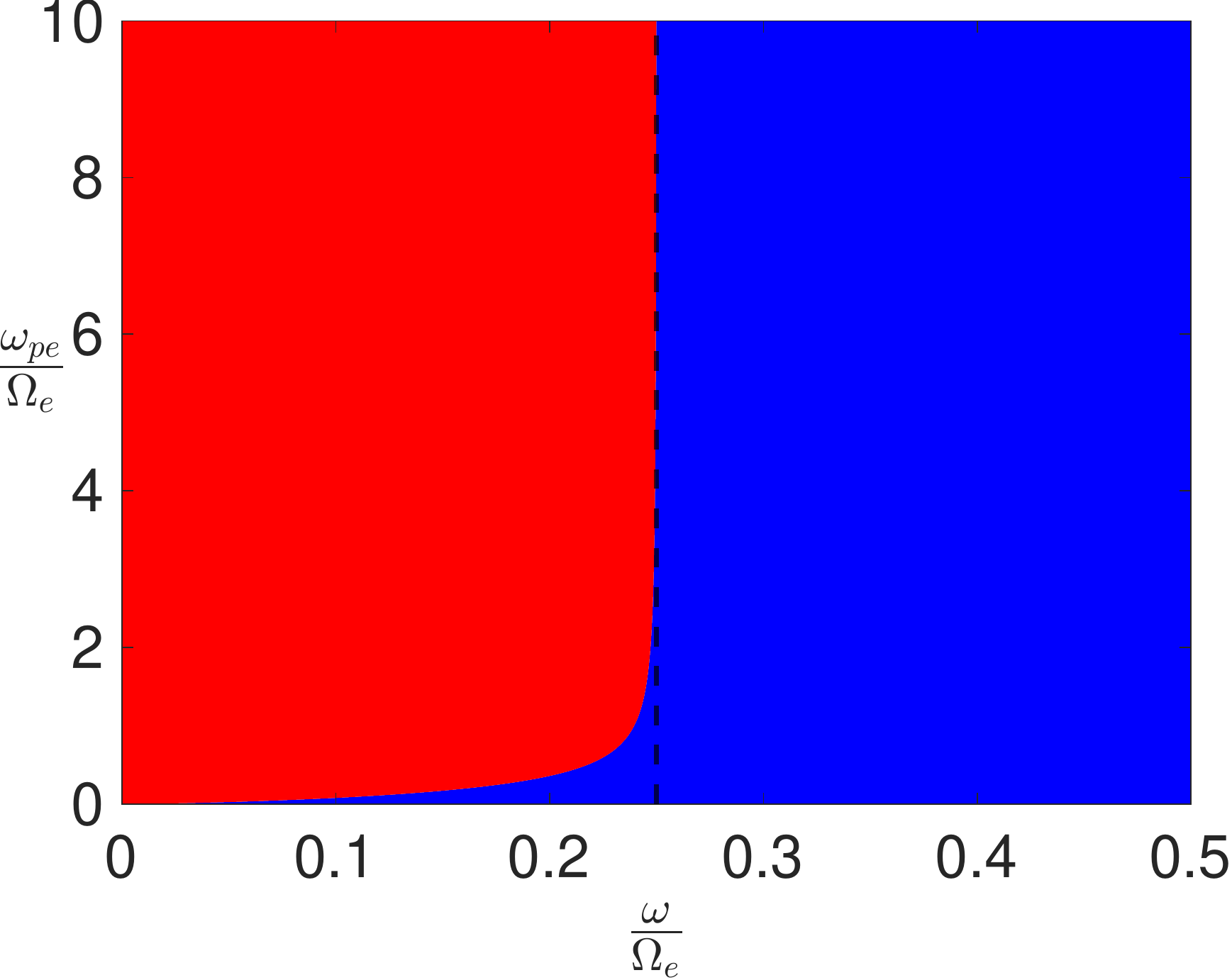}
    \end{subfigure}
        \begin{subfigure}{0.4\textwidth}
    \includegraphics[width=\textwidth]{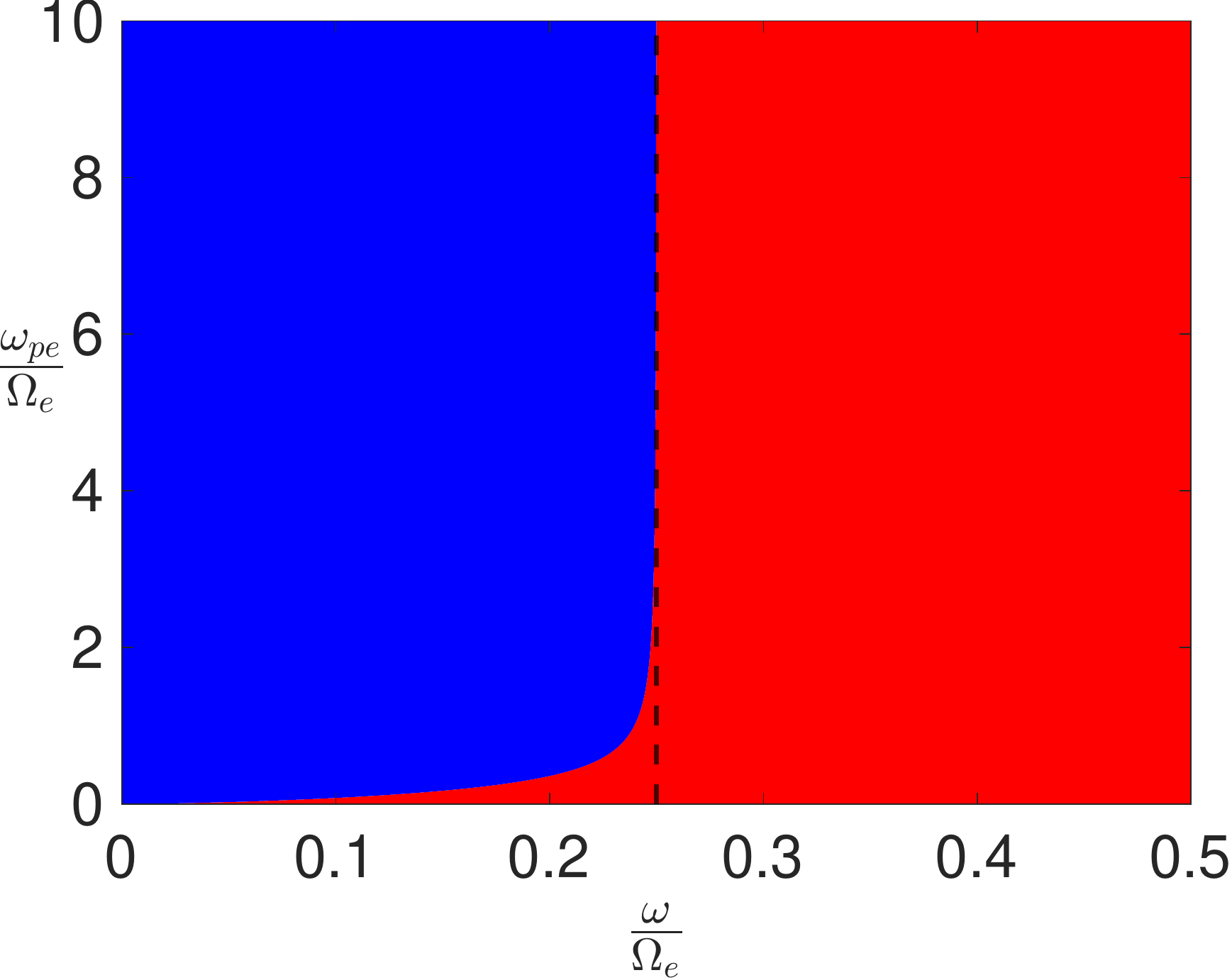}
    \end{subfigure}
    \caption{\centering Signs of the right hand side term in (blue) and instability (red) for parallel propagating Whistler waves for subsonic (left) and supersonic (right) waves. Dashed line marks the asymptote $\omega = \frac{\Omega_e}{4}$}
    \label{fig:stability}
\end{figure}

These modulation equations for the wave parameters can now be analysed for their stability, which is achieved by investigating small perturbations to some fixed state. We take this constant state to be ${\bf U}_0 = (k_0,\mathscr{B}_0,0,N_0)$, noting that the choice of velocity does not meaningfully impact the results that follow. By considering perturbations of the form ${\bf U}={\bf U}_0+\delta \hat{U}e^{i(Z-CT)}$, then the leading order perturbation is governed by the eigenvalue problem
\[
({\bf A}-C{\bf I})\hat{\bf U} = {\bf 0}\,.
\]
The resulting eigenvalues $C$ ultimately determine the stability of this system - if they are all real the constant state ${\bf U}_0$ is stable and the monochromatic wave perseveres, however when any of these eigenvalues is complex (occurring in complex conjugate pairs) there is an exponentially growing mode that causes the perturbation to rapidly diverge from the monochromatic wave state~\citep{wlnlw}. The emergence of these complex eigenvalues is more commonly referred to as a \emph{modulational instability}. The characteristic polynomial for this problem admits 4 roots in general, which can be categorised by their values as $\mathscr{B}_0\,N_0 \to 0$:
\[
C = \pm c_s\,, \quad C = \frac{\partial\omega_0}{\partial k} \equiv c_g \ \mbox{(multiplicity 2)}.
\]
We focus on the latter set of roots, as it transpires that the eigenvalues associated with the sound speed can be shown to be real but those associated with the group velocity can become complex. This is typical of problems involving waves coupled to a mean field, where it is the wave mode driving the instability\citep{Bridges-2022,Tam-1970}, and so is not unexpected here either. For this problem, these latter roots can be expanded in powers of $\mathscr{B}_0$ and $N_0$, again assumed small, to give
\[
C = c_g\pm \sqrt{\omega_0''\omega_2\mathscr{B}_0}+\mathcal{O}(\mathscr{B}_0,N)
\]
where the effective nonlinear frequency correction $\omega_2$ is given by
\[
\omega_2 = \frac{\omega_{pe}^4\omega^2( \Omega_e-\omega)(\Omega_e\omega_{pe}^2 + 2\Omega_e\omega^2 - 2\omega^3)}{c^2(c_g^2-c_s^2)(2\Omega_e^2\omega + \Omega_e\omega_{pe}^2 - 4\Omega_e\omega^2 + 2\omega^3)^2)}\,.
\]
It is clear that these eigenvalues are complex whenever $\omega_0''\omega_2<0$. To determine when this occurs more readily, we introduce the nondimensional scalings
\begin{equation}\label{nondim}
\omega = \Omega_e W\,, \quad k = \frac{\Omega_e}{c}K\,, \quad c_g = cV\,, \quad \omega_{pe} = \alpha \Omega_e\,, \quad c_s = c \nu\,,
\end{equation}
giving
\[
\begin{split}
\omega_0'' &= \frac{2c^2\alpha^2(W - 1)^2(4W^4 - 4W^3 - 4W\alpha^2 + \alpha^2)}{\Omega_e(2W^3 - 4W^2 + \alpha^2 + 2W)^3}\,, \\[3mm]
\omega_2 &= \frac{\alpha^4\Omega_e^4W^2(W - 1)(2W^3 - 2W^2 - \alpha^2)}{c^4(V^2-\nu^2)(2W^3 - 4W^2 + \alpha^2 + 2W)^2}
\end{split}
\]
The only sign changes that happen within the Whistler wave interval $0<W<1$ are the root of $\omega_0''$,
\[
\frac{W^3(W-1)}{W-\frac{1}{4}} = \alpha^2 \,,
\]
which for large $\alpha$ typical in the earth's radiation belts approaches $W = \frac{1}{4}$, a quarter of the gyrofrequency. The other sign change is due to the factor $V^2-\nu^2$ passing through zero, which occurs when the Whistler wave's group velocity goes from subsonic ($c_g<c_s$) to supersonic ($c_g>c_s$). Thus we have the following criterion for the modulation stability of Whistler waves:
\begin{equation}
    \mbox{Modulational instability when}
    \begin{cases}
        W^3(W-1)<\alpha^2(W-\frac{1}{4})\,, \quad |V|>\nu \quad ({\rm supersonic})\\[3mm]
        W^3(W-1)>\alpha^2(W-\frac{1}{4})\,, \quad |V|<\nu \quad ({\rm subsonic})
    \end{cases}
\end{equation}
This information is summarised in figure \ref{fig:stability}. It should be noted there are cases where $|V|\sim \nu$ where the story is partially more complex and introduces a further stability boundary, but this is not generic and will not be discussed in detail here. As the speed of sound scales linearly with temperature within the setting considered, we can infer that the supersonic case is more prevalent in cold plasmas, with the subsonic case being expected in warmer plasmas. However, some caution should be noted here as the isothermal approximation is known to be poorer for warmer plasmas \citep{Chen-2012,Gao-2014,Li-2010b}, so a more technical theory to confidently conclude anything regarding this limit.

In summary, we have used classical modulation theory to explore the stability of monochromatic WMC waves with the additional consideration of the particle effects. Criterion for modulation instability, associated with the formation of wave packets, is deduced by exploring the nature of the eigenvalues of the $4\times 4$ quasilinear system (\ref{mod-system}), revealing that both the group velocity dispersion $\omega_0''$ and the difference between the squares of the group and sound speed $c_g^2-c_s^2$. This identifies where one should look for the subelement structures typical of WMC waves, and the analysis of their evolution forms the remainder of this paper. To do so we will rely on the classical perturbative approach for the evolution of the wave envelope which we derive in the next section.

\section{Ion Effects and Packet Generation}
\label{sec:WNT}
The classical Whitham modulation approach of the previous section grants us insight into the criteria necessary for WMC waves to develop into packets. It is the case, however, that to formally derive an equation for the spatiotemporal evolution of these packets one must also include the effects of ions within the analysis. It is not \emph{a priori} obvious that this is necessary, especially as their role is greatly overshadowed by effects due to electron motion, but this analysis will highlight that ionic features, particularly the number fraction of ions present in the plasma, has a crucial role in the modulation stability of WMC.

With ion motions included, we will consider the two-fluid plasma description~\citep{baumjohann-2012} with the assumption of isothermality for each particle species. This gives the system of equations
\begin{gather}
    \nabla \times \be = -\pd{\bb}{t}\,,\label{mhd-eq-1}\\
    \nabla \times \bb  = \mu_0 (q_e n_e\bve+q_in_i \bvi) +\frac{1}{c^2}\pd{\be}{t}\,,\label{mhd-eq-2}\\
    \pd{\bve}{t}+(\bve \cdot \nabla)\bve +\frac{c_{s,e}^2}{n_e}\nabla n_e= \frac{q_e}{m_e}\left(\be+\bve \times \bb\right)+{\bf F}_{P,e}\,,\label{mhd-eq-3}\\
    \pd{\bvi}{t}+(\bvi \cdot \nabla)\bvi +\frac{c_{s,i}^2}{n_i}\nabla n_i= \frac{q_i}{m_i}\left(\be+\bvi \times \bb\right)+{\bf F}_{P,i}\,,\label{mhd-eq-4}\\
    \pd{n_e}{t}+\nabla \cdot (n_e\bve) = 0\label{mhd-eq-5}\\
    \pd{n_i}{t}+\nabla \cdot (n_i\bvi) = 0\label{mhd-eq-6}\\
    \nabla \cdot {\bf E} = q_en_e+q_in_i\,. \label{mhd-eq-7}
\end{gather}
where the subscripts $i,\,e$ denote fields and quantities which describe the ion and electron populations respectively. The ponderomotive forces for each species will be taken as
\[
{\bf F}_{P,j} = -\frac{\omega_{pj}^2}{2 \mu m_j n_j c^2 \omega^2}\nabla \langle |{\bf E}|^2\rangle)\,, \quad j = e,\,i\,.
\]
The ponderomotive force for the ion equation is of the same form as the electron, which can be obtain by following the derivation in \citet{cipp}. As in the purely electron case, alternate forms of this force may be supplied instead but this primary form of the ponderomotive force captures the essence of the wave-particle interaction.
For the spatiotemporal analysis, we will undertake a formal weakly nonlinear analysis in the small parameter $\eps \ll 1$. This parameter is a characterisation of the wave amplitude, which in turn determines the strength of the nonlinear effects. Following typical amplitude equation approaches, we postulate the following expansions:
\[
\begin{split}
\bb &= B_0\kh+\eps(\ih+i\jh)\big(B_W(Z,T)+\eps\beta_1(B_W)_Z\big) e^{i\theta}\\[3mm]
\be &= \eps(\ih+i\jh)\big(\alpha_1B_W(Z,T)+\eps \beta_2(B_W)_Z\big)e^{i\theta}\\[3mm]
\bve &=v_{||}\kh + \eps(\ih+i\jh)\big(\alpha_2B_W(Z,T)+\eps \beta_3(B_W)_Z\big)e^{i\theta}+\eps^2V_e(Z,T)\kh\\[3mm]
\bvi &=u_{||}\kh + \eps(\ih+i\jh)\big(\alpha_3B_W(Z,T)+\eps \beta_4(B_W)_Z\big)e^{i\theta}+\eps^2V_i(Z,T)\kh\\[3mm]
n_e &= n_{e,0}+\eps^2N_e(Z,T)\,, \qquad n_i = n_{i,0}+\eps^2N_i(Z,T)\,, \qquad \theta = kz-\omega t \\
\end{split}
\]
where the slow variables $Z$ and $T$, encoding two time scales, are defined as
\[
Z = \eps(z-c_gt)\,, \quad T = \eps^2 t\,,
\]
whose scalings are chosen so that the terms of the derived evolution equation are all of the same order in $\eps$. 
The smallness of the parameter $\eps$ is once again determined by the separation between linear and nonlinear scales, and following a similar calculation as that in \S 2, one finds that $\eps$ must satisfy the ordering

\[
\eps  \left\vert\frac{B_W}{B_0}\right\vert\sin \phi \ll \left \vert\frac{\omega -v_{||}k}{\Omega}\right \vert
\]
which for field aligned WMC with a beam velocity parallel to the magnetic field is automatic, as $\sin \phi =0$.
Such scalings are typical within weakly nonlinear theories describing spatiotemporal amplitude evolution and ultimately indicate that the evolution equation for the amplitude is of Nonlinear Schr\"odinger type.
Indeed, substitution of the above expansions into the governing equations and solving the resulting problems at each order of $\eps$ confirm this, as we will outline below.

At leading order, $\mathcal{O}(\eps)$, we find that $\omega,k$ satisfy the relation
\[
\begin{split}
D = \frac{1}{c^2}\bigg[ &(c^2k^2-\omega^2)(\omega-v_{||}k-\Omega_e)(\omega-u_{||}k+\Omega_i)\\
& \qquad +\omega_{pe}^2(\omega-v_{||}k)(\omega-u_{||}k+\Omega_i)+\omega_{pi}^2(\omega-u_{||}k)(\omega-v_{||}k-\Omega_e)\bigg] = 0
\end{split}
\]
which has a root corresponding to the Whistler dispersion relation
\[
c^2k^2-\omega^2+\frac{\omega_{pe}^2(\omega-v_{||}k)}{\omega-v_{||} k-\Omega_e} +\frac{\omega_{pi}^2(\omega-u_{||}k)}{\omega-u_{||} k+\Omega_i} = 0
\]
Waves along this dispersion branch require that
\[
(\alpha_1,\alpha_2, \alpha_3) = \left(\frac{ i \omega}{k},\,-\frac{q_e}{m_e k}\frac{\omega-v_{||}k}{\omega-v_{||}k-\Omega_e},\,
-\frac{q_i}{m_i k}\frac{\omega-u_{||}k}{\omega-u_{||}k+\Omega_i}\right)
\]
The next order of the analysis generates both first harmonic and zero harmonic terms. The former of these may be solved to show that $\beta_1 = 0$ and
\[
\begin{split}
(\beta_2,\beta_3,\beta_4)= 
\left(
\frac{c_gk-\omega}{k},\,
-\frac{i q_e ((\omega-v_{||}k)^2-\Omega_e(\omega-c_g k) )}{m_e k^2(\omega-v_{||}k - \Omega_e)^2},\,
-\frac{i q_i ((\omega-u_{||}k)^2+\Omega_i(\omega-c_g k) )}{m_i k^2(\omega-u_{||}k + \Omega_i)^2}
\right)
\end{split}
\]
The first of the zero harmonic (i.e. oscillation-free) problems arises from (\ref{mhd-eq-2}) and is simply
\begin{equation}\label{zero-current}
q_e(n_{e,0}V_e+N_ev_{||})+q_i(n_{i,0}V_i+N_iu_{||}) = 0
\end{equation}
This is to say that the parallel propagating wave does not induce an additional current at this order of the analysis. The second arises from (\ref{mhd-eq-7}) and gives that
\begin{equation}\label{quasineutral}
q_eN_e+q_iN_i = 0\,,
\end{equation}
which is equivalent to quasineutrality holding in the presence of slow deviations. This allows one to write the electron number density variations according to
\[
N_e = -\frac{q_i}{q_e}N_i = \mathcal{Z}_cN_i\,,
\]
where $\mathcal{Z}_c$ is the ion charge number.
Both of the conditions (\ref{zero-current}) and (\ref{quasineutral}) highlight the importance of ionic effects, as without the ionic contribution this equation would necessarily yield that $N_e,\,V_e$ must vanish and lead to no nonlinear effects emerging from the weakly nonlinear analysis. Thus this order makes it clear that ionic effects must be included in the study of the multiple scale WMC evolution and should not be neglected.

The problem at $\mathcal{O}(\eps^3)$ is where the analysis terminates, and only the first harmonic and zero harmonic terms need to be considered to develop the evolution equation that results here. The zero harmonic terms are more involved at this order, and these read
\begin{equation}\label{zero-harm-final}
\begin{split}
(v_{||}-c_g)N_e'+n_{e,0}V_e' &= 0\,,\\
(u_{||}-c_g)N_i'+n_{i,0}V_i' &= 0\,,\\
(v_{||}-c_g)V_e'+\frac{c^2_{s,e}}{n_{e,0}}N_e' &=- \frac{\omega_{pe}^2}{\mu m_en_{e,0} c^2 k^2}(|B_w|^2)'\\
(u_{||}-c_g)V_i'+\frac{c^2_{s,i}}{n_{i,0}}N_i' &= -\frac{\omega_{pi}^2}{\mu m_i n_{i,0} c^2 k^2}(|B_w|^2)'\,,
\end{split}
\end{equation}
where primes denote derivatives with respect to $Z$.
The equation (\ref{zero-current}) to show that is we add $q_e$ times the first equation to $q_i$ times the second, we necessarily have that
\[
c_g(q_eN_e+q_iN_i)' = 0\,, \quad \implies \quad
q_eN_e+q_iN_i = {\rm constant}.
\]
From (\ref{quasineutral}) we can see that it follows that this constant is zero. We will recast the system (\ref{zero-harm-final}) by taking the third equation of (\ref{zero-harm-final}) and subtracting $\frac{q_i}{q_e} = -\mathcal{Z}_c$ times the fourth. Overall, this gives that
\[
\begin{split}
\mathcal{Z}_c(v_{||}-c_g)N_i'+n_{e,0}V_e' &= 0\,,\\
(u_{||}-c_g)N_i'+n_{i,0}V_i' &= 0\,,\\
(v_{||}-c_g)V_e'+Z(u_{||}-c_g)V_i'+\mathcal{Z}_c\bigg(\frac{c_{s,e}^2}{n_{e,0}}+\frac{c_{s,i}^2}{n_{i,0}}\bigg)N_i' &=-\frac{Z\omega_{pi}^2m_en_{e,0}+\omega_{pe}^2m_in_{i,0}}{\mu m_em_i n_{i,0} n_{e,0} c^2 k^2}(|B_w|^2)'
\end{split}
\]
This system of equations can be inverted to show that the modulation of the mean particle quantities are related to the carrier wave by
\begin{equation}\label{bul-mode-results}
\begin{split}
&\begin{pmatrix}
N_i\\
V_e\\
V_i
\end{pmatrix} = 
\frac{\mathcal{Z}_c\omega_{pi}^2m_en_{e,0}+\omega_{pe}^2m_in_{i,0}}{\mathcal{Z}_c\mu m_em_in_{e,0}n_{i,0} c^2 k^2\Delta}(|B_w|^2)
\begin{pmatrix}
n_{e,0}n_{i,0}\\
\mathcal{Z}_c(c_g-v_{||})n_{i,0}\\
(c_g-u_{||})n_{e,0}\\
\end{pmatrix}\,, \\[3mm]
{\rm with} \qquad &\Delta = \big((c_g-u_{||})^2-c_{s,i}^2\big)n_{e,0}+\big((c_g-v_{||})^2-c_{s,e}^2\big)n_{i,0}\,.
\end{split}
\end{equation}
The arbitrary functions resulting from the integration have been ignored here, as these simply correspond to a shift in the frequency of the carrier wave which evolves much more slowly than the envelope.

At $\mathcal{O}(\eps^3)$, we require that all terms proportional to the first harmonic vanish, else the analysis will generate secular terms. It can be shown that in order to do so, the amplitude $B_W$ must satisfy the following Nonlinear Schr\"odinger (NLS) equation
\begin{equation}\label{NLS}
    iB_T+\frac{\omega_0''}{2}B_{ZZ}-\Gamma |B|^2B = 0\,,
\end{equation}
where the nonlinear frequency correction $\Gamma$ is given by
\begin{multline}
\Gamma = \frac{\mathcal{Z}_c\omega_{pi}^2m_en_{e,0}+\omega_{pe}^2m_in_{i,0}}{\mathcal{Z}_c\mu c^4m_em_in_{e,0}n_{i,0} k^2\Delta D_\omega}(\omega-v_{||}k-\Omega_e)(\omega-u_{||}k+\Omega_i)\\
\times \bigg(\frac{\mathcal{Z}_c\omega_{pe}^2 n_{i,0}}{(\omega-v_{||}k-\Omega_e)^2}\left((\omega-v_{||}k)^2-\Omega_e  (\omega-c_gk)\right)\\
+\frac{\omega_{pi}^2 n_{e,0}}{(\omega-u_{||}k+\Omega_i)^2}\left((\omega-u_{||}k)^2+\Omega_i(\omega-c_gk)\right)\bigg)
\end{multline}

We conclude our formal derivation with a few marks about the validity and limitations of the above model. Primarily we expect the NLS equation above to be a good representation whenever the variations in the wave amplitude occur over scales much larger than the wave period, which is to say that the model is representative whenever the typical packet contains many waves. Observation from the Van Allen probes and THEMIS mission support this being typical of WMC~\citep{Zhang-2019,Artemyev-2022c}. It is also known that such envelope models are valid for evolution times up to $\mathcal{O}(\eps^{-2})$, so one can expect much longer predictions for lower amplitude packets. However, nonlinear events in WMC typically happen on scales of less than a second~\citep{santolik-2008}, with the most repetitive emissions taking place within windows of several seconds~\citep{Gao-2022}, suggesting that these events lie within the timespan of model validity. Finally a consequence of the scalings of the moving co-ordinate we have a narrow spectrum requirement of $|k-\delta k|,\,|\omega- \delta \omega| \sim \mathcal{O}(\eps)$, where $\delta k,\,\delta \omega $ represent the wavenumber and frequency associated with the amplitude $B_W$, a constraint that mission data suggests WMC satisfies~\citep{santolik-2003,santolik-2008}.

\subsection{Influence of Ions on Modulation Stability}
In order to analyse the modulation instability of Whistler mode waves, we will again nondimensionalise the wave quantities according to (\ref{nondim}), and introduce the further nondimensionalisations
\begin{equation}\label{params}
r = \frac{m_e}{m_i}\,, \quad c_{s,e} = c \nu\,.
\end{equation}
Additionally, to simplify the analysis in the first instance, we will chose that $v_{||} = u_{||} = 0$.
We note that the parameter $r$ is typically small, with its largest value occurring for hydrogen ions where it takes the value $r = \frac{1}{1836}$. Thus we treat $r \ll 1$. This simplifies the coefficients of the NLS (\ref{NLS}) to
\[
\begin{split}
\omega_0'' &= \frac{2\alpha^2c^2}{\Omega}\frac{(W - 1)^2(4W^4 - 4W^3 - 4W\alpha^2 + \alpha^2)}{(2W^3 - 4W^2 + \alpha^2 + 2W)^3}+\mathcal{O}(r)\\[3mm]
\Gamma &= \frac{\alpha^4 n_{i,0} \Omega}{\mu m_en_{e,0} \Delta}\frac{(W - 1)^2(2W^3 - 2W^2 - \alpha^2)}{ (2W^3 - 4W^2 + \alpha^2 + 2W)^2(W^2 - \alpha^2 - W)}+\mathcal{O}(r)\,,\\[5mm]
{\rm with} \quad  \Delta &= c^2(n_{e,0}+n_{i,0})\left[V^2-\frac{n_{i,0}}{n_{e,0}+n_{i,0}}\nu^2\right]+\mathcal{O}(r)\,.
\end{split}
\]
It becomes clear that although the vast majority of effects due to ions characterised by $r$ vanish, more or less recovering the modulation instability criterion of the electron-only plasma presented in \S \ref{sec:MI-electron}. There is, however, a non-negligible ion effect that is crucial in the expression for $\Delta$ associated with the subsonic to supersonic wave transition. The remaining expression involving the ion number density now controls one of the stability boundaries for the Whistler waves, as was demonstrated in \S \ref{sec:Mod1}. This expression indicates that the proportion of ions present in the plasma directly alter this boundary, and so the super- and subsonic regimes are altered to transition at ratios that can be much lower than unity. This is since
\[
0\leq \frac{n_{i,0}}{n_{e,0}+n_{i,0}} \equiv \eta_i \leq 1\,.
\]
Further, in the limit as $\eta_i\to 0$ where the electron fluid (i.e. static electrons) description is expected to be employed, one finds that the sound speed term is eliminated and only the supersonic case is operational, suggesting that parallel WMC would only rise in tone and parallel falling tones are expected to be non-existent in this regime. This is in strong agreement with the mission data on parallel WMC waves, where rising tones are closely aligned with the magnetic field whereas falling tones are to be expected nearly orthogonal to it \citep{Taubenschuss-2014,Teng-2019}. We make this inference primarily based on the location of the region within which the wave is stable, however dynamically speaking we do not have a term which dictates the movement of the main spectral peak in frequency space. The nonlinear Schrodinger will need to be extended in order to accommodate such terms and test part of this hypothesis, which is precisely the aim of the next section of our analysis.

\section{Spectral Asymmetry and the Emergence of Chirp}
\label{sec:chirp}
It is known that NLS models do not admit chirping behaviour for monochromatic waves or wave packets (see, for example, the standard textbooks \citet{Ablowitz-1981,Billingham-2000,Agrawal-2000} on the subject). This owes to the fact that the growth of sidebands due to modulation instability are symmetric and thus it has no preferred spectral shift. This limits the effect of the NLS solutions on the original carrier wave to simply a constant, time-independent frequency shift. To break this spectral symmetry, it is necessary to appeal to higher order effects within the weakly nonlinear theory that induce self-frequency shifting as has been explored in optics \citep{Palacios-1999,Goyal-2011,triki-2016,Triki-2022}.

We return our discussion to the approach outlined in \S \ref{sec:WNT}. To obtain these higher order effects, it is necessary to analyse the contributions from the first harmonic terms at $\mathcal{O}(\eps)$. The terms these generate are then included into the NLS equation (\ref{NLS}) and treated as correction terms (as these are, strictly speaking, of a lower order than the original terms). The result of doing this for parallel propagating Whistlers is the modified NLS equation
\begin{equation}\label{MNLS}
iB_T+\frac{\omega_0''}{2}B_{ZZ}-\Gamma |B|^2B-i\eps Q |B|^2B_Z = 0
\end{equation}
where the additional nonlinear correction has coefficient
\begin{multline}
Q = \frac{Z\omega_{pi}^2m_en_{e,0}+\omega_{pe}^2m_in_{i,0}}{Z\mu m_em_in_{e,0}n_{i,0} c^4 k^3\Delta D_\omega}(\omega-v_{||}k-\Omega_e)(\omega-u_{||}k+\Omega_i)\\
\times\left[\frac{Z\omega_{pe}^2n_{i,0}}{(\omega-v_{||}k-\Omega_e)^2}\left((\omega-v_{||}k)^2-\Omega_e(\omega-c_gk)\right)\left(1+\frac{(c_g-v_{||})k}{\omega-v_{||}k-\Omega_e}\right)\right.\\
\left.+\frac{\omega_{pi}^2n_{e,0}}{(\omega-u_{||}k+\Omega_i)^2}\left((\omega-u_{||}k)^2+\Omega_i(\omega-c_gk)\right)\left(1+\frac{(c_g-u_{||})k}{\omega+\Omega_i}\right)\right]
\end{multline}

We now demonstrate how this additional term alters the growth of sidebands due to a modulation instability. This may be investigated by perturbing the Stokes wave solution, an exact monochromatic wavetrain solution to this amplitude equation, by
\[
B = A(1+P)e^{i [\delta k Z- (\omega_0'' \delta k^2/2+ A^2(\Gamma-\eps Q \delta k) T]}\,,
\]
for Stokes wave amplitude $A$, sideband wavenumber $\delta k$ and $P$ is a perturbation assumed small enough that quadratic terms in it are negligible. Upon substitution, it can be shown that the perturbation $P$ admits the linear spectrum
\[
\left(\sigma -\eps \kappa QA^2\right)^2 = \kappa^2\bigg[\omega_0''(\Gamma-\eps Q \delta k) +\frac{\omega_0''^2}{4}\kappa^2\bigg]
\]
and thus there is a band of spectral wavenumbers $\kappa$ which generates perturbations that grow in time whenever
\[
\omega_0''\Gamma \left(1-\frac{\eps Q \delta k}{\Gamma}\right) <0\,,
\]
We can see that the modulation criterion $\omega_0''\Gamma<0$ may persist if the bracketed expression is positive. This is ensured even if $\delta k>0$ as for parameter ranges operational in the magnetosphere one has $\eps Q/\Gamma\ll1$, and so $\delta k$ must be significantly large in order to do so which would violate the assumptions used to derive (\ref{MNLS}). Thus the modulational instability has exactly the same thresholds as the NLS equation.
The asymmetry in the wavenumber spectrum can then be determined by studying the maximum growth rate for the perturbed Stokes wave. This can be calculated as
\[
G = {\rm max(Im}(\sigma)) = |\Gamma+\eps Q \delta k|\,, \quad {\rm at} \quad \kappa = \pm \sqrt{-\frac{2(\Gamma-\eps Q \delta k)}{\omega_0''}}
\]
It is apparent now that the sideband wavenumber $\delta k$ introduces bias to one side of the spectrum via this higher order term characterised by $Q$. This is determined by the sign of $Q$, which like $\Gamma$ is determined by the factor $\Delta$. Thus, terms with a lower sideband wavenumber will grow at a faster rate than their upper sideband counterpart. What happens subsequently cannot be inferred by this linear stability analysis, and requires further nonlinear approaches that we will not consider here.

This observation is the first hint that this higher order term will be the driver of the chirping behaviour that we seek to understand. We will continue to explore this extended NLS equation from the perspective of nonlinear wave packets, the structures observed within WMC waves that ultimately generate the rising and falling tones we are interested in understanding. These structures will be derived and explored analytically and numerically in later parts of this section to explore how these may encourage a similar spectral asymmetry in WMC waves.

\subsection{Emergence of the Band Gap}
Of significant note within this weakly nonlinear theory, when accounting for the higher order terms, is that the classical band gap at half the gyrofrequency is a natural consequence. We can identify this by using the nondimensionalisations (\ref{nondim}) and (\ref{params}), where $r \ll 1$, to simplify the coefficient of the term responsible for chirping to
\[
Q = \frac{\alpha^4cn_i}{\mu m_e n_{e,0} \Delta  K}\frac{(W - 1)^2(2W - 1)(2W^2 - \alpha^2 - 2W)(2W^3 - 2W^2 - \alpha^2)}{ (W^2 - \alpha^2 - W)(2W^3 - 4W^2 + \alpha^2 + 2W)^3}
\]
In this limit there is clearly a root of $Q$ at $W = \frac{1}{2}$, namely $\omega = \Omega_e/2$, which is at the typical band-gap for WMC waves. The only other sign change of $Q$ is determined by $\Delta$ (which ultimately leads to a breakdown of the weakly nonlinear theory as a whole) as the remaining factors result in a positive definite quantity in the lower band WMC range $0<W\leq\frac{1}{2}$.

This observation of this root of $Q$ at $W=\frac{1}{2}$ is an important one. It demonstrates that at half the gyrofrequency the amplitude evolution equations once again reduce to the NLS equation, so the wave mode neither biases a rise or fall in the frequency. Thus there is no emergent sweep rate for waves at this frequency, and it is clear that waves with frequencies close to this band gap will have very weak sweep rates too. Overall, this observation suggests that a WMC element that rise or fall in frequency will have its sweep rate continuously reduced as its spectral peak approaches this band gap until the wave becomes incoherent or its energy is exchanged with another wave of a different frequency. 

Unfortunately, the narrow-band picture of the nonlinear theory is insufficient to fully test this hypothesis. For energy exchanges between different frequencies and how the wave responds, a multi-mode interaction system is required in the spirit of either a Coupled NLS/Manakov system~\citep{manakov-1974,kourakis-2005,kourakis-2005b,baronio-2012} or a Zakharov/Wave Turbulence model~\citep{Galtier-2000,newell-2011,david-2022}, both much more complex dynamical models. Therefore a full validation of this continual sweep rate reduction postulation is reserved for later study. We however note that the former of these extensions is essentially a coupled system of NLS-like equations and will retain many of the coefficients derived in this paper, and thus the argument regarding an arrested sweep rate in the proximity of the band gap we hypothesise to hold true.

\subsection{Wave Packets with Chirp}
Now that we have a modification to the NLS, as well as some insight into how chirp might come about for sidebands, let us formally derive a solution for a Chorus element that undergoes chirping. To do so, we follow~\cite{di93} and consider a solution of (\ref{MNLS}) of the form
\[
B = R(\xi)e^{i (\nu Z-\sigma T)}e^{i \phi(\xi)}\,, \quad \xi = Z-VT\,.
\]
which is formed of three parts - the amplitude function $R$, the sideband wave component represented by the second term and a phase function $\phi$, which will ultimately be the source of chirp. This is because the local frequency of this solution can be defined as the negative of time derivative of the total phase function,
\[
\tilde{\omega} = \frac{\partial}{\partial t}(\sigma T-\nu Z-\phi) = \eps(c_g\nu+\eps\sigma) +\eps(c_g+\eps V)\phi'\,.
\]
From this one may identify a sweep rate, defined as the rate of change of the frequency in time:
\[
\mathscr{S} = \frac{\partial \tilde{\omega}}{\partial t} = -\eps^2(c_g+\eps V)^2\phi''
\]
Our aim to to relate this sweep rate with properties of the wave packet solution to determine the resulting frequency change, which is achieved though substitution of this ansatz into (\ref{MNLS}) and solving the resulting ODEs.

Starting with our guess at a solution, and splitting the resultant system into real and imaginary parts, we have the equations
\begin{gather}
    \sigma R+V\phi'R+\frac{\omega_0''}{2}(R''-(\nu+\phi')^2R)-\Gamma R^3+\eps Q (\nu+\phi')R^3 = 0 \label{re-part}\\
    -VR'+\frac{\omega_0''}{2} (2R'\phi'+R\phi''+2 \nu R')-\eps Q R^2R' = 0\label{im-part}\,.
\end{gather}
We start by taking the second equation, (\ref{im-part}), and multiplying by $R$, which makes it an exact derivative. Integrating gives
\begin{equation}\label{int-eqtn}
-\frac{V}{2}R^2+\frac{\omega_0''}{2} (R^2\phi'-\nu R^2)-\frac{\eps Q}{4}R^4 = I
\end{equation}
for some constant $I$. It will be conventient herein to introduce $U = R^2>0$, and note that for solitary wave packets one must have that $U,U'$ and $U''$ tend to zero as $\xi \to \pm \infty$. This yields that $I=0$ and allows one to manipulate (\ref{int-eqtn}) to show that the sweep rate $\mathscr{S}$ is
\begin{equation}\label{sweep-rate}
\mathscr{S}  = -\frac{\eps^3(c_g+\eps V)^2 Q}{2\omega_0''}U'
\end{equation}
In modulationally unstable regions, this prefactor is negative, meaning that a positive sweep rate occurs when $U'>0$, i.e. at points where the envelope of an element is increasing.

We conclude the discussion here with an exact solution describing the envelope of a single Whistler packet/element. To do so, we note that we can set $\nu = 0$ to simplify the analysis, and it can be reintroduced under suitable mappings. In this case, we get a quartic potential:
\[
(U')^2+B_2 U^2+B_3U^3+B_4U^4 \equiv (U')^2-\mathscr{V}(U) = 0
\]
with
\[
B_2 =\frac{4V^2}{\omega_0''^2}+\frac{8\sigma}{\omega_0''} \,, \quad B_3 =\frac{4}{\omega_0''}\left(\frac{\eps VQ}{\omega_0''}-\Gamma\right) \quad B_4 = \left(\frac{\eps Q}{\omega_0''}\right)^2\,.
\]
A homoclinic connection ( that is a trajectory that begins and terminates at the same fixed point) within this system, corresponding to a solitary wave solution is only possible so long as there is an interval for which $\mathscr{V}(U)>0$ . As $B_4>0$, we need $B_2<0$ and so
\[
\frac{4V^2}{\omega_0''^2}+\frac{8\sigma}{\omega_0''} <0
\]
Since $\sigma$ is a free real-valued parameter, this is always possible to satisfy. Following~\citep{k12} let us factorise this ODE as follows:
\begin{equation}\label{Wave-ode}
U' =\left\vert\frac{\eps Q}{\omega_0''}\right\vert\sqrt{U^2(U-U_-)(U_+-U)}\,,
\end{equation}
where the roots are given by
\[
U_\pm =\frac{2\omega_0''}{\eps^2Q^2}\left[\Gamma-\frac{\eps V Q}{ \omega_0''} \pm \sqrt{\Gamma^2-\frac{2\eps Q}{\omega_0''}(V\Gamma+\eps Q \sigma)} \ \right]
\]
The negative subscript denotes the negative root and plus the positive root so that $U_-<0<U_+$.
Thus, the expression in (\ref{Wave-ode}) under the square root is positive in the interval $U_-<U<U_+$, but as $ U = R^2>0$ the only interval of interest to us will be $0\leq U<U_+$. The solution to this ODE is of the form of a Gardner/extended KdV soliton \citep{k12,Grimshaw-2010}:
\[
U = \frac{A}{(1+B)\cosh^2 \Theta - 1}
\]
with
\[
A = -U_->0\,, \quad B = -\frac{U_-}{U_+}>0\,, \quad \Theta = \sqrt{-\frac{2 \sigma \omega_0''+V^2}{\omega_0''^2}}(Z-VT)
\]
and so the amplitude of this wave packet is $A/B = U_+$.

In summary, by considering higher order terms in the amplitude model we find that these terms are the source of chirping behaviour in both monochromatic wavetrains and within wave packets. In the case of the latter, however, we can observe that there is no net shift in the frequency due to the symmetry of the packet. Chirping due to WMC elements would appear to arise, therefore, upon packet interaction as this leads to asymmetry within the wave envelope and thus generates a net frequency shift. This is now explored using numerical means, where we will generate multiple packets, observe their interactions and extract their frequency spectrum to deduce the overall chirp produced by their interplay.

\section{Numerical Simulation of Interacting Whistler Mode Elements}
\label{sec:numerics}
It is clear from the available spacecraft mission data that Whistler and Chorus wave packets are rarely isolated and propagate in groups of multiple packets. As a result, these packets interact and cause changes to their frequency and amplitude that the above insight of a solitary packet cannot provide. To investigate these interactions, we resort solving (\ref{MNLS}) using a timestepping procedure. For computational ease and to better identify the effects of system parameters on the sweep rate, we rescale (\ref{MNLS}), highlighting two cases dependant of the sign of $\omega_0''$. We employ the scalings
\[
X = {\rm sign}(\omega_0'') Z\,, \quad \tau = {\rm sign}(\omega_0'') 
\frac{\omega_0''}{2}T \,, \quad  B = 
\begin{cases}
    \sqrt{\frac{|\omega_0''|}{2|\Gamma|}}A^* & \omega_0''<0 \\
    \sqrt{\frac{|\omega_0''|}{2|\Gamma|}}A & \omega_0''>0 
\end{cases}
\]
to transform (\ref{MNLS}) into
\begin{equation}\label{nondim-MNLS}
iA_\tau+A_{XX}-\sigma\left(|A|^2A+\frac{i\eps Q}{\Gamma}|A|^2A_X\right) = 0\,, \quad {\rm with} \quad \sigma = {\rm sign}(\omega_0''\Gamma)\,.
\end{equation}
This reduces the problem to a single tunable parameter,
which we can from the nondimensionalisations earlier that 
\begin{equation}\label{ratio}
\frac{\eps Q}{\Gamma} = \frac{\eps c}{\Omega K}\frac{(2W-1)(2W^2-2W-\alpha^2)}{2W^3-4W^2+2W+\alpha^2}\,,
\end{equation}
thus reducing the tunable parameters to just $\eps,\,W,\,\Omega$ (or equivalently, $B_0$) and $\alpha$ which are determined by the plasma environment. For the majority of our simulations we will be using parameter choices representative of those found at $L$-shell $L\sim 6$. This corresponds to
\[
\alpha = 7.2\,, \quad B_0 =1.4\times 10^{-7} \,, \quad n_e = 1\times 10^7\,, \quad \eps = \mathcal{O}(10^{-5}).
\]
In addition to this, we focus on wave frequencies lying in the lower frequency band range of $0.1 \Omega $-$0.5\Omega$.

We advance (\ref{nondim-MNLS}) in time using an exponential time differencing method with fourth order Runge-Kutta timestepping (ETDRK4) \citep{cm02,Kassam-2005}, using periodic boundary conditions to take advantage of the speed and spectral accuracy of Fourier-based schemes. 
To initialise a multi-element solution and remain close to the analytic solution found, we initialise the simulations with the wave packet
\[
A(X,0) = \sqrt{\frac{1}{A_0\cosh\left(\frac{X}{\Lambda}\right)^2-1}} e^{i p X}\,,
\]
which has amplitude $\frac{1}{A_0-1}$, width $\Lambda$ and sideband wavenumber $p$. The width of these has to be suitably large (of the order $10$) in order to generate multiple packets, with smaller $\Lambda$ recovering a single wave packet with small amplitude wave radiation and larger widths generating many packets that partially fission before interaction. In our simulations, we allow this initial profile to evolve and we then observe the dynamics of this fissioned structure in time to determine how the emergent wave packets interact with each other and the resulting Fourier spectrum.

\begin{figure}
    \centering
    \begin{subfigure}{0.8\textwidth}
            \includegraphics[width=\textwidth]{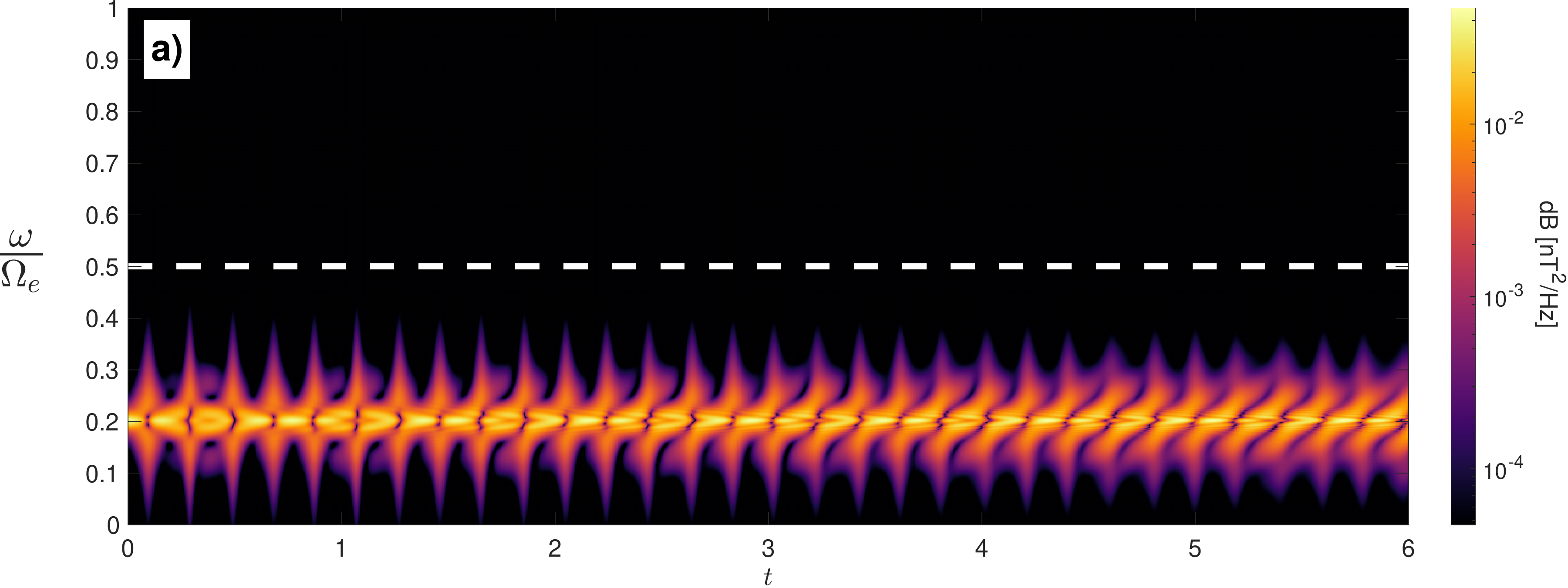}
    \end{subfigure}
        \begin{subfigure}{0.8\textwidth}
            \includegraphics[width=\textwidth]{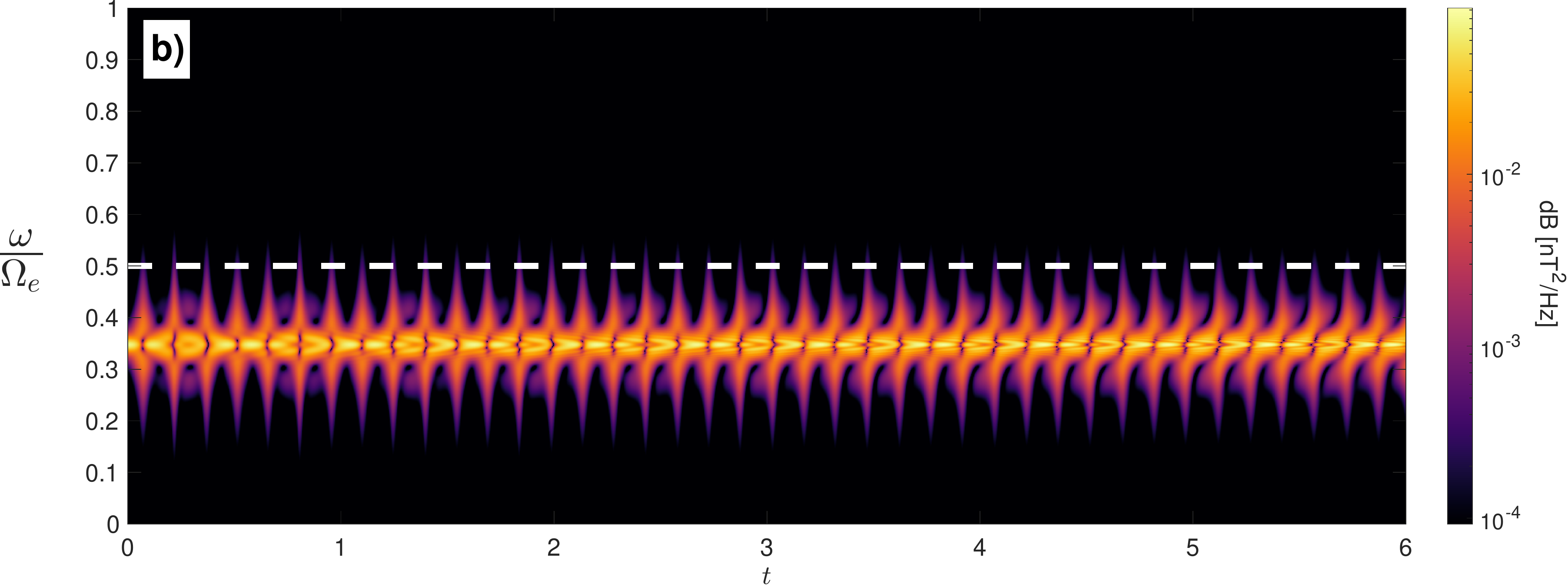}
    \end{subfigure}
        \caption{Examples of the power spectrum $|B_W|^2/\delta \omega$ generated by the parameter choices $(\eps,n_e,\alpha,\Omega) = (4\times 10^{-5},1\times 10^7,7.2,2.49\times 10^4)$ and a) $W=0.2,\, k_BT =8.02$KeV, b) $W=0.35,\, k_BT =7.37$KeV.}
    \label{fig-powerspec}
\end{figure}

The result of simulating (\ref{nondim-MNLS}) yields wave packets that generate frequency sweeps, with examples appearing in figure \ref{fig-powerspec}. These emerge after the initial packet begins to split and interact with the new subelements, and develop further as the packets separate. This separation drives an envelope asymmetry, which from our observations in (\ref{sweep-rate}) would appear to the source for the frequency sweeps that emerge numerically. The sweep rate decreases over each simulation, as the subelements grow further apart, further reinforcing that it is the packet interactions that are the source of rising tones. We find over the course of our numerical investigations that higher magnitude of the ratio in (\ref{ratio}) enhance the onset of frequency sweeping, which further points to the role of the modified term in the NLS plays in generating frequency shifts in the WMC waves.

\begin{figure}
    \centering
\includegraphics[width=\textwidth]{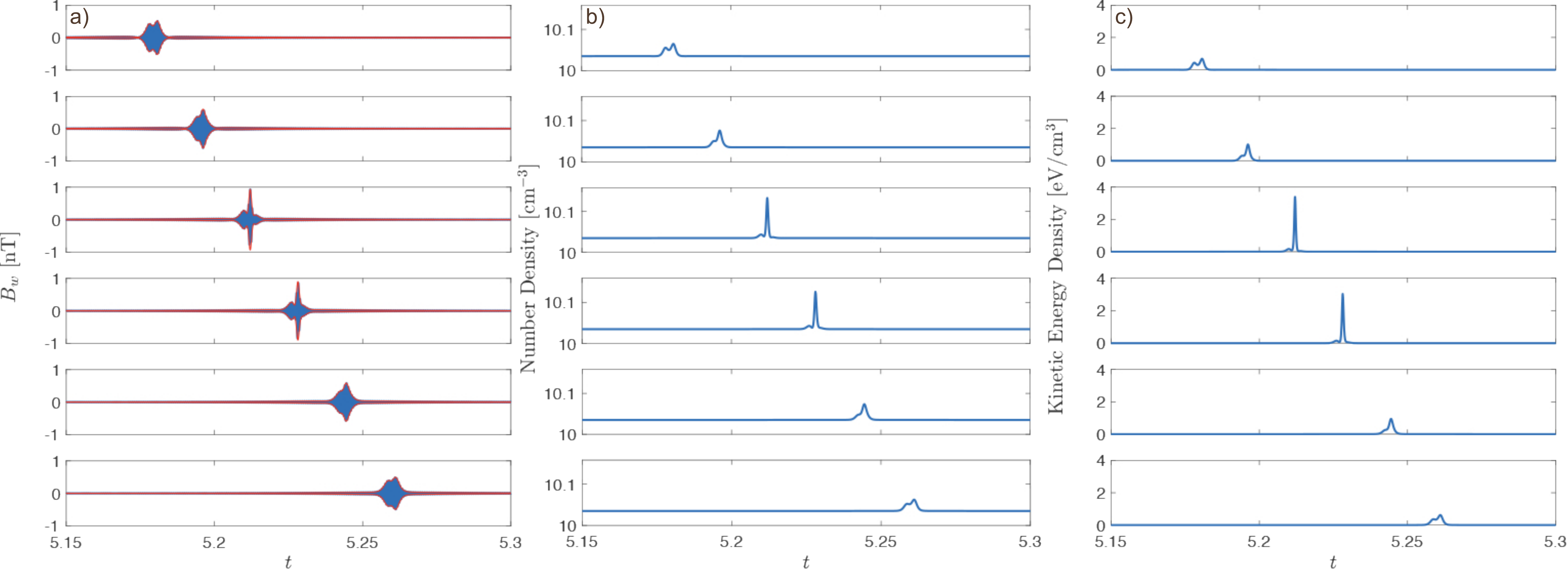}
    \caption{Snapshots of the time series of a) the magnetic wave b) number density and c) kinetic energy density (right) at several spatial points, demonstrating the breather-like evolution of the wave envelope as the WMC wave travels.}
    \label{fig:profiles}
\end{figure}

\newpage

The structure of the envelope generated by the interaction of subelements is also noteworthy. In our simulations we find that the packets exhibit an almost time-periodic breathing behaviour, depicted in figure \ref{fig:profiles}a) akin to the Kuznetsov–Ma soliton found in the NLS \citep{Ma-1979,Akhmediev-1985}, albeit with a zero background. The structure of maximum amplification for widths that generate 2-3 main packets also bears some semblance to higher order rogue wave solutions \citep{Chabchoub-2012,Slunyaev-2013}, a fact that is not unsurprising given that such solutions are the infinite-period limit of breathers. This seems to suggest, therefore, that isolated rising-tone WMC waves may fall into the category of rogue waves themselves, and the repetitive WMC emission events are breather events. This connection between hydrodynamic rogue waves and WMC has not yet been made in the literature and it may be useful to explore this connection further in the future, given also that the occurrence rate of large-amplitude WMC is significant in the magnetosphere.

\begin{figure}
    \centering
    \begin{subfigure}{0.74\textwidth}
            \includegraphics[width=\textwidth]{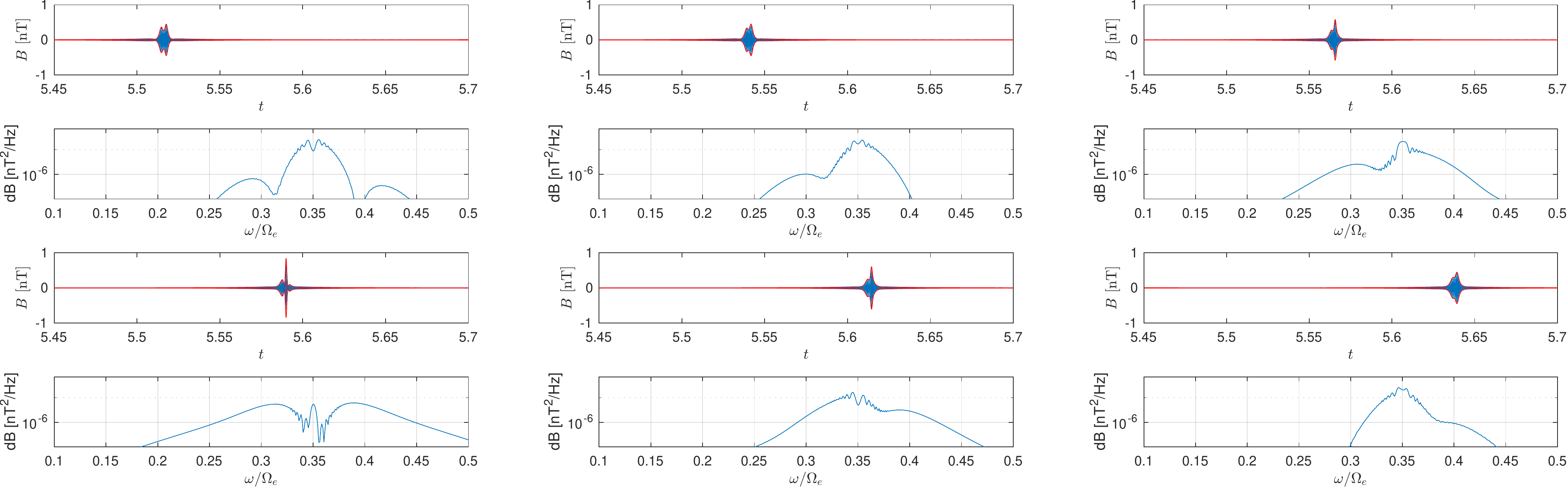}
            \end{subfigure}
    \begin{subfigure}{0.1725\textwidth}
            \includegraphics[width=\textwidth]{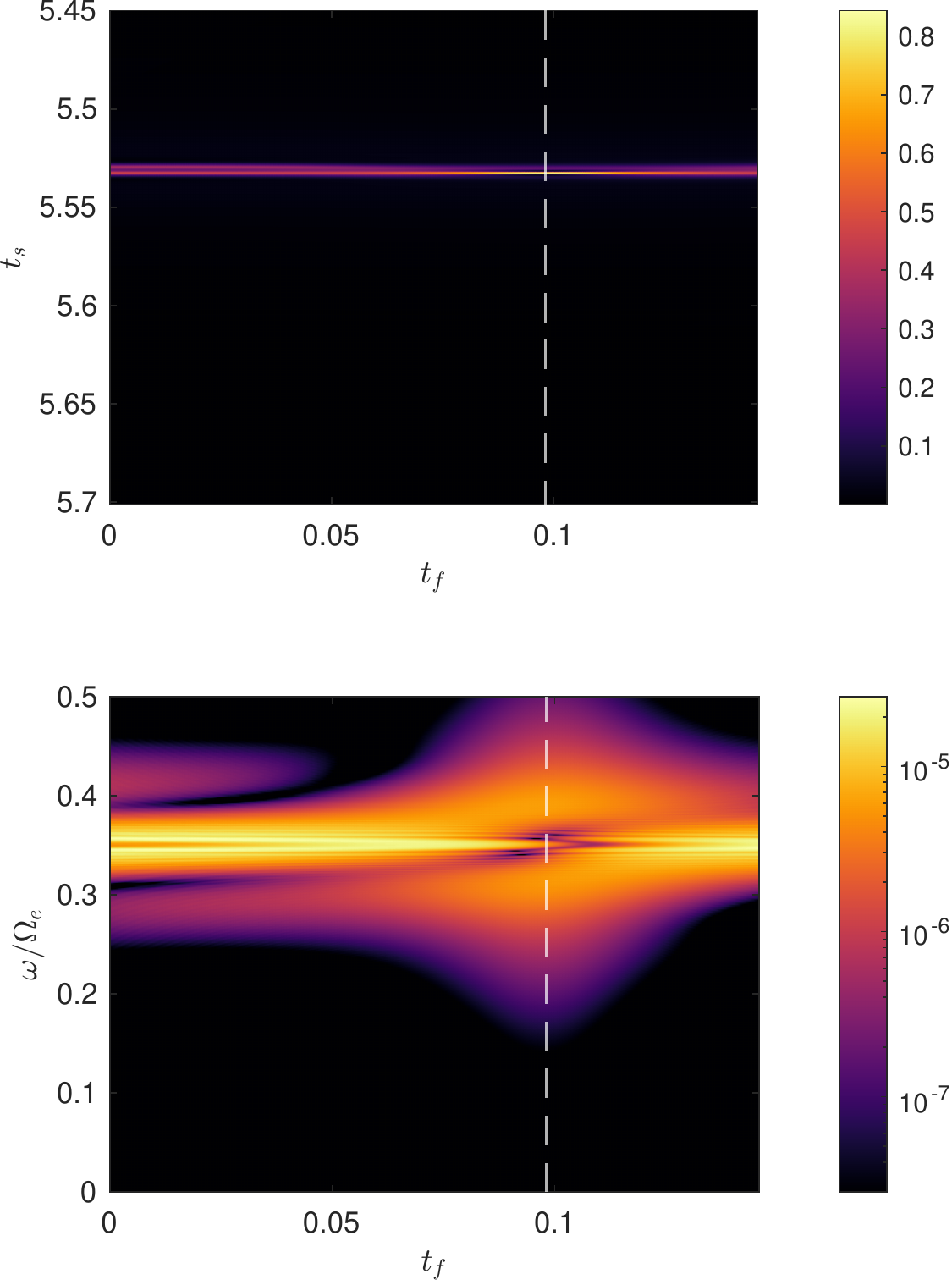}
            \end{subfigure}
        \caption{Left: Snapshots of the magnetic field wave versus the power spectra for the time series snapshot for the parameter choice $(\eps,n_e,\alpha,\Omega, W, k_BT) = (4\times 10^{-5},1\times 10^7,7.2,2.49\times 10^4,0.35, 7.373$ KeV). Right: Comparison between the wave envelope (with maxima shifted to the same point in slow time $t_s$) over the time frame of one pulsation (the fast time $t_f$) and its short time Fourier transform. The white dashed line denotes the time at which the envelope achieves its maximum.}
    \label{fig:climb}
\end{figure}

We may also use (\ref{bul-mode-results}) to explore the effect on both the number density and the electron beam velocity. From the latter, we may also extract the following energy density, a combination of kinetic and thermal energy of the electrons induced by the wave:
\[
E_{KE} = \frac{m_e}{2}n_e||{\bf V}_e-v_{||}||^2+\frac{k_BT}{2}(n_e-n_{e,0}),
\]
which allows us to determine which stage of the WMC evolution energises the electrons. We visualise these quantities in figure \ref{fig:profiles}c), demonstrating that the maximum amplification of the wave is the stage which imparts the most energy to the particles, increasing the energy by several electronvolts. We may also correlate this moment with the frequency activity of the wave, as we do in figure \ref{fig:climb}. In it, we can observe that the amplification of the wave develops as one rising tone begins to terminate and another, lower in frequency, rising tone begins to initiates. The kinetic energy density peaks at exactly the point in time where the power of the higher frequency rising tone and the lower frequency rising tone are equal, and beyond this point the lower rising tone takes over as the most powerful part of the signal. This observation would therefore suggest that repetitive rising tone events, that are typical of chorus wave activity \citep{Tsurutani-1974,Li-2012,Gao-2014}, are accelerating electrons most when the tones overlap and there is an exchange in wave power between these elements. It is difficult to conclusively identify this as the amplification mechanism however, as it could simply be that if the rising tone of higher frequency have been left to climb the amplification may have been much greater. Further study beyond that of this paper will be required to fully explore this interaction and its consequences.

We conclude this section with some final commentary on what aspects of WMC we cannot recreate with our current level of modelling, but could be captured with suitable alterations. Primarily, the MNLS equation appears to be unable to produce a single isolated rising tone event that has been successfully created via other theoretical modelling (for example, \citet{Nunn-1997,Omura-2008,Tao-2014b}). We attribute this to the fact that we have no source term corresponding to a hot electron population which ultimately drives the WMC waves in these existing theories. We anticipate that once such effects are accounted for properly, the simulations here should be able to reproduce these results. Further, another aspect of our simulations we are not able to reproduce is the non-sequential gap between events, as all WMC events in our simulations happen one after another. One can see from the mission data (cf. \cite{Agapitov-2011,Gao-2022}) that this need not be the case, and there can be considerable separation between WMC occurrences. We hypothesise that once hot electron growth effects are accounted for, these too may be emergent from simulations of our amplitude model.

\section{Concluding Remarks}
\label{sec:conc}

This paper has provided an overview of the formation and dynamics of parallel propagating WMC wave with rising tone. We have outlined the mechanism for packet formation is an instability of modulational type, whose transition is marked by a critical point in the group velocity as well as the sign of $c_g^2-c_s^2$. Further investigations demonstrated that the role of ions within the system is to reduce the latter instability threshold and suggests that their effects, although contributing little elsewhere, are non-negligible and must be accounted for in any analysis of WMC.

We have observed here that rising tones emerge from the NLS with higher order effects, and the degree at which the spectral peak changes can be attributed to either the strength of the background magnetic field  and/or the frequency of the original carrier wave. We have also numerically investigated the evolution of multiple WMC wave packets, confirming that the interaction between WMC elements seems to be the mechanism for the emergence of their frequency sweeping. The envelope asymmetry appears to be the reason for this as this interplay overcomes the symmetry of the theoretical solitary WMC element solution which seemingly does not allow for. This suggested mechanism is supported by the data from the THEMIS and Van Allen missions, where the wave profile for rising tone elements contains both multiple elements and asymmetric wave envelopes.

The assumption of narrow-banded waves implicit in the derivation of the envelope equations captures the phenomenology of WMC. We do, however, note that formally the range of frequencies that the packets can sweep is limited by $\eps$ and should not cover a broad range of frequencies simultaneously. This means that the complete picture of a rising tone WMC will involve more complex models which couple multiple frequency bands together and transfer wave energy between one another. This has been seen in other systems, the simplest being coupled NLS models with the limit of such couplings being the Zakharov equation. For parallel propagating WMC, the theoretical analysis should remain tractable due to the lack of second harmonic terms in the perturbation analysis and yield further insight into the stability and evolution of nonlinear Chorus waves.

An important extension of the approach outlined here is to consider the more general case of WMC propagating obliquely to the magnetic field. This is more reflective of what is observed, where WMC is seen to most commonly propagate at angles of $10^\circ$ (close to parallel) or $70^\circ$ to the magnetic field. The work here remains reflective of the former of these cases by virtue of the obliqueness remaining small, but the latter case would require a revised version of the perturbative approach and result in a different modified NLS. The mission data would also suggest that this version of the dynamics would instead lead to wave dynamics which admit falling tone WMC, a feature which may be linked to a negative version of the ratio (\ref{ratio}). In either case, a quantification of the effect of oblique propagation should be determined and explored in a similar fashion to the field-aligned waves considered in this paper.

\section*{Acknowledgments}
Daniel Ratliff is grateful to the Isaac Newton Institute for Mathematical Sciences, Cambridge, for support and hospitality during the programme Dispersive Hydrodynamics where work on this paper was undertaken. This work was supported by EPSRC grant no EP/R014604/1.

Oliver Allanson gratefully acknowledges financial support from the University of Exeter, the University of Birmingham, and also from the United Kingdom Research and Innovation (UKRI) Natural Environment Research Council (NERC) Independent Research Fellowship NE/V013963/1.

\appendix

\section{Details of  the WKB analysis of the electron plasma}
\label{app:WMT}
First order terms proportional to $e^{i \theta}$ are given by the linear matrix problem
\begin{equation}\label{det-sys}
\begin{pmatrix}
i \omega & k&0\\
-k&\frac{i \omega}{c^2} & \mu_0 q n_0\\
-\frac{iq v_{||}}{m}&-\frac{q}{m}&i(\omega - v_{||}k-\Omega_e)
\end{pmatrix}
\begin{pmatrix}
1\\
\alpha_1\\
\alpha_2
\end{pmatrix} \equiv {\bf D}(\omega,k;v_{||})
\begin{pmatrix}
1\\
\alpha_1\\
\alpha_2
\end{pmatrix} =  {\bf 0}\,.
\end{equation}
The determinant of this matrix needs to be zero for nontrivial solutions and thus
\[
D_W(\omega,k;v_{||}) = |{\bf D}| =  \frac{1}{c^2}\bigg[(\omega-v_{||}k-\Omega_e)(c^2k^2-\omega^2)+\omega_{pe}^2(\omega-v_{||}k)\bigg] = 0
\]
Useful for later are the following derivative results along the branch of solutions:
\[
\begin{split}
(D_W)_\omega &= -\frac{1}{c^2}\bigg[\frac{\omega_{pe}^2\Omega_e}{\omega-v_{||}k-\Omega_e}+2\omega(\omega-v_{||}k-\Omega_e)\bigg]\\
(D_W)_k &= \frac{1}{c^2}\bigg[2c^2k(\omega-v_{||}k-\Omega_e) + \frac{v_{||}\omega_{pe}^2\Omega_2}{\omega-v_{||}k-\Omega_e}\bigg]\,,\\
(D_W)_{v_{||}} &= \frac{1}{c^2}\,\frac{\omega_{pe}^2\Omega_e k}{\omega-v_{||}k-\Omega_e}\\
(D_W)_{n_0} &=\frac{\omega_{pe}^2(\omega-v_{||}k)}{n_0c^2}\,.
\end{split}
\]
For Whistler waves, one has that
\[
\alpha_1 = -\frac{i \omega}{k}\,, \qquad \alpha_2 = -\frac{q}{mk}\frac{\omega-v_{||}k}{\omega-v_{||}k-\Omega_e}\,.
\]
The left eigenvector of this matrix is also required in order to determine criterion for when
\[
{\bf D}{\bf A}_1 = {\bf A}_2
\]
can be solved for ${\bf A}_1$, namely that the Fredholme alternative is to hold and the right hand side vanishes when projected in the direction of the left eigenvector. This is given by
\[
{\bf l} = \left( \frac{\omega(\omega-v_{||}k-\Omega_e)-\omega_{pe}^2}{c^2k},i(\omega -v_{||}k-\Omega_e),-\mu_0 q n_0\right)
\]
One is then able to show that
\[
{\bf l}{\bf D}\begin{pmatrix}
1\\\alpha_1\\\alpha_2
\end{pmatrix} = -\frac{i D}{k}\,.
\]

The next order of the analysis only requires consideration of zero harmonic modes, associated to wave-particle interactions. This involves terms in which can be written in terms of gradients, and so by considering the expressions under the gradients we have the system
\[
\begin{split}
&q \mu (v_{||}N+n_0V) = 0\,,\\
\nabla \left( v_{||}V+\frac{c_s^2}{n_0}N+\frac{\omega_{pe}^2}{\mu c^2 \omega^2}|\alpha_1|^2 |B_W|^2 \right) = &\nabla \left( v_{||}V+\frac{c_s^2}{n_0}N+\frac{\omega_{pe}^2}{\mu c^2 k^2} |B_W|^2 \right) = 0\,.
\end{split}
\]
Combining these gives
\[
\frac{c_s^2-v_{||}^2}{n_0}N+\frac{\omega_{pe}^2}{\mu c^2 k^2} |B_W|^2 = {\rm Constant}
\]
This constant can be deduced by noticing that the wave-free flow is potential,as there is no indication that such waves have vorticity due to the constant underlying background magnetic field, so that
\[
\bve = \nabla \phi = \nabla (v_{||}z-\gamma t)\,.
\]
This would mean that the overall version of this bulk equation would be
\[
\frac{c_s^2-v_{||}^2}{n_0}N+\frac{\omega_{pe}^2}{\mu c^2 k^2} |B_W|^2 = \gamma -\frac{v_{||}^2}{2}-c_s^2\ln n_0
\]

At the final order we consider, we must now examine the terms proportional to $e^{i \theta}$, given by
\begin{equation}\label{final-ord}
{\bf 0} = \begin{pmatrix}
0\\
-q \mu \alpha_2 N B_W\\
\frac{iq}{m}B_wV+i k V\alpha_2 B_w
\end{pmatrix}
\end{equation}
Now, we project the right hand side using ${\bf l}$ gives
\[
\begin{split}
\frac{i}{c^2 k}\bigg[\frac{\omega_{pe}^2 \Omega_e k}{\omega - v_{||} k -\Omega_e}\, V+\frac{\omega_{pe}^2(\omega - v_{||}k)}{n_0}N \bigg]B_w   = \frac{i\omega_{pe}^2}{c^2kn_0}\frac{(\omega-v_{||}k)^2-\Omega_e \omega}{\omega-v_{||}k-\Omega_e}NB_w
\end{split}
\]
Now, we could impose that this is zero, but the analysis would be determined to be trivial. Instead, we assume that there is a correction to the frequency, $\delta \omega$, that has amplitude effects within it. Assuming $\omega  = \omega_0+\delta\omega$, the matrix system (\ref{det-sys}) has the amplitude dependent part
\[
\delta \omega 
\begin{pmatrix}
i&0&0\\
0&\frac{i}{c^2}&0\\
0&0&i
\end{pmatrix}
\begin{pmatrix}
1\\
\alpha_1\\
\alpha_2
\end{pmatrix}B_W = i\delta \omega B_W\begin{pmatrix}
1\\\frac{\alpha_1}{c^2}\\ \alpha_2
\end{pmatrix}
\]
which would instead give the equation (\ref{final-ord}) as
\[
i\delta \omega B_W\begin{pmatrix}
1\\\frac{\alpha_1}{c^2}\\ \alpha_2
\end{pmatrix}=  \begin{pmatrix}
0\\
-q \mu \alpha_2 N B_W\\
\frac{iq}{m}B_wV+i k V\alpha_2 B_w
\end{pmatrix}
\]
Projecting this down to assess whether this is solvable gives
\[
-i \delta \omega \frac{(D_W)_\omega}{k} = \frac{i\omega_{pe}^2}{c^2kn_0}\frac{(\omega-v_{||}k)^2-\Omega_e \omega}{\omega-v_{||}k-\Omega_e}N
\]
and so
\[
\delta\omega = -\frac{i\omega_{pe}^2 }{(D_W)_\omega c^2n_0}\frac{(\omega-v_{||}k)^2-\Omega_e \omega}{\omega-v_{||}k-\Omega_e}N\,.
\]
Thus, the two equations that emerge from the WKB theory are
\[
\begin{split}
D B_w+\frac{\omega_{pe}^2}{c^2n_0}\frac{(\omega-v_{||}k)^2-\Omega_e \omega}{\omega-v_{||}k-\Omega_e}NB_W = 0\,,\\
\frac{c_s^2-v_{||}^2}{n_0}N+\frac{\omega_{pe}^2}{\mu c^2 k^2} |B_W|^2 = \gamma -\frac{v_{||}^2}{2}-c_s^2\ln n_0\,.
\end{split}
\]
\bibliographystyle{jpp}

\bibliography{referados}

\end{document}